\documentstyle[12pt]{article}
\title{\huge Anisotropic effect on two-dimensional cellular automaton traffic flow with
periodic and open boundaries} 
\author{\bf  
 A. Benyoussef, H. Chakib and H. Ez-Zahraouy%
\thanks{Corresponding author : ezahamid@fsr.ac.ma} 
\\
 Laboratoire de Magn\'{e}tisme et de la Physique
 des Hautes Energies
\\
Universit\'{e} Mohammed V, Facult\'{e} des Sciences, Avenue Ibn Batouta,  B.P. 1014
\\Rabat, Morocco
}
\date{ }

\begin{document}
\maketitle 

\abstract{By the use of computer simulations we investigate, in the cellular automaton of
two-dimensional traffic flow, the
anisotropic effect of the probabilities of the change of the move directions of cars, from up to right ($p_{ur}$) and from right to up ($p_{ru}$), on the dynamical
jamming transition and velocities under the periodic boundary
conditions in one hand and the phase diagram under the open
boundary conditions in the other hand. However, in the former case, the first order jamming transition
disappears when the cars alter their directions of move ( $p_{ur}\neq 0$ and/or $p_{ru}\neq 0$). In the
open boundary conditions, it is found that the 
 first order line transition between jamming and
moving phases is curved. Hence, by increasing the anisotropy, the
moving phase region expand as well as the contraction of the
jamming phase one. Moreover, in the isotropic case, and  when  each car changes its direction of move every time steps ($p_{ru}=p_{ur}=1$), the transition from the jamming phase (or moving phase) to
the maximal current one is of first order. Furthermore, the density
profile decays, in the maximal current phase, with an exponent
$\gamma \approx \frac{1}{4}$.}

\newpage

\section{\protect\bigskip Introduction}

Transport phenomena in complex systems, in particular models of highway
traffic flow, attracted much attention in recent years. Much of the effort
was concentrated on discrete stochastic models of traffic flow, first
proposed by Nagel and Schrekenberg \cite{1}, and subsequently studied by
many other authors using a variety of techniques \cite{2}-\cite{5}. Since
the introduction of the Nagel Shrekenberg (NaSch) model \cite{1}, cellular
automata became a well established method of traffic flow modeling.
Comparatively low computational cost of cellular automata models made it
possible to conduct large-scale real-time simulations of urban traffic in
the city of Duisburg \cite{6} and Dallas/Forth Worth \cite{7}. Compared with
the fluid dynamical approaches to traffic flow problems, the CA models are
conceptually simpler, and can be readily implemented on computers. These
models have the advantages that they can be easily modified to deal with the
effects of different kinds of realistic conditions, such as road blocks and
hindrances, traffic accident \cite{8}, highway junctions \cite{9}, vehicle
acceleration \cite{10}, stochastic delay due to drivers reactions \cite{5},
anisotropy of car distributions in different driving directions \cite{11},
faulty traffic lights \cite{12}. Traffic flow is a kind of many body systems
of strongly interacting cars. Recent studies reveal physical phenomena such
as the dynamical phase transitions and nonlinear waves \cite{13},\cite{14}.
When the car density increases, the jamming transition occurs and traffic
jams appear. The jamming transitions from the freely moving traffic to the
jammed traffic have been studied by microscopic and macroscopic models. The
two-dimensional traffic flow is more complex than the one-dimensional case.
The two-dimensional traffic has been investigated only by the cellular
automaton models \cite{15}-\cite{18}. It has been shown that the jamming
transition occurs in the two-dimensional case and is similar to the
one-dimensional case.

The NaSch model \cite{1} is a probabilistic CA model for one-dimensional
highway traffic. It considered the effects of acceleration and stochastic
delay of vehicles with high speed. A vehicle can move at most $v_{\max }$
sites in a time step, where $v_{\max }$ is the maximal velocity. The speed
at a time step depends on the gap (number of empty sites) between successive
vehicles. If the speed in the present time step is less than $v_{\max }$ and
the gap ahead allows, the speed increases by one unit in the next time step.
If the spacing ahead is less than the speed in the present time step, then
the speed is reduced to the value allowed by the spacing. The speed of a car
is reduced by one unit in the next time step with a probability $p$,
exhibiting a randomization in realistic traffic flows. We introduced in a
recent article \cite{19} the boundary effects on the NaSch model, a vehicle
can enter without constraint, with a probability $\alpha $, in the first
site being to the left side of the road if this site is empty. While, a
vehicle being on the right in the last site can leave the road with a
probability $\beta $. There is a free flow and jamming phase separated by a
line of first order transitions, this transition occurs at $\alpha <\beta $
for $p\neq 0$ and $\alpha =\beta $ for $p=0$, and the maximal current phase
is obtained only for $v_{\max }=1$ (this case coincides with the asymmetric
exclusion process model), otherwise it vanishes. Cheybani et al. \cite{20}
introduced a constraint during the entry of a vehicle, where at site $i=0$,
that means out of the system a vehicle with the probability $\alpha $ and
with the velocity $v=v_{\max }$ is created. This car immediately moves
according to the NaSch rules. If the velocity of the injected vehicle on $%
i=0 $ is $v=0$, then the injected vehicle is deleted. For $p$ greater than a
critical value, the maximal current phase separated by second order
transitions occurs like for the asymmetric simple exclusion process (ASEP).

The two-dimensional models have been presented by Biham, Middeleton and
Levine (BML) to mimic the traffic flow in the whole city. The BML model \cite
{15} is a simple two-dimensional (square lattice) CA model. Each cell of the
lattice represents an intersection of an east-bound and a north-bound
street. The spatial extension of the streets between two intersections is
completely neglected. The cells (intersections) can either be empty or
occupied by a vehicle moving to the east or to the north. In order to enable
movement in two different directions, east-bound vehicles are updated at
every odd discrete time-step whereas north-bound vehicles are updated at
every even time-step. The velocity update of the cars is realized following
the ASEP rules, a vehicle moves forward by one cell if the cell in front of
is empty, otherwise the vehicle does not move. The alternating movement to a
traffic lights cycle of one time-step. The traffic-flow model is given by a
three-state CA on the square lattice. Biham, Middeleton and Levine have
studied the traffic-flow problem only in the case $\rho _{x}=\rho _{y}=\frac{%
\rho }{2}$ where $\rho _{x}$ and $\rho _{y}$ are respectively the density of
cars moving to the right and the density of cars moving upwards, and $\rho $
is the total density of cars. They have found that a dynamical jamming
transition occurs at a critical density $\rho =\rho _{c}$ with increasing
the density of cars. The dynamical jamming transition separates between the
low-density moving phase and the high-density jamming phase. Nagatani \cite
{16} has investigated the anisotropic effect of density of cars on BML
models, it was shown that the traffic-jam transition occurs at higher
density of cars with increasing the difference between the density of right
moving cars and the density of up moving cars. The difference of the
densities of cars has an important effect on the dynamical jamming
transition. Cuesta et al. \cite{21} introduced the randomness
parameter $\gamma $ which allows to control the trend of the motion of every
car. Half of cars move horizontally (vertically) with the probability $%
\gamma $ ($1-\gamma $) (accordingly, $1-\gamma $ ($\gamma $) is the
probability to move vertically (horizontally) ). The system in this case
exhibits the phase diagram of a first order phase transition from a freely
moving to a jammed phase. The curves presenting the variation of the mean
velocity versus the density $\left( \rho \right) $ undergo a discontinuous
transition of magnitude $\Delta v\left( \gamma \right) $ at the transition $%
\rho _{c}$. As $\gamma $ increases, $\rho _{c}$ increases and $\Delta
v\left( \gamma \right) $ decreases, and eventually vanishing for some
randomness $\gamma _{c}$.
Our aim in this paper is to study, using numerical simulations, the effect of the anisotropy of the probabilities of the change of the move directions of cars, from up to right ($p_{ur}$) and from right to up ($p_{ru}$),in the cellular automaton of two-dimensional traffic flow, on the dynamical jamming transition and velocities under the periodic boundary conditions in one hand and the phase diagram and density profile behaviour under the open boundary conditions in the other hand. 

The paper is organized as follows; in the following section we define the
model, the section 3 is reserved for results and discussions, the conclusion
is given in section 4.

\section{Model}

We exhibit a simpler cellular automaton model that describes traffic flow in
two dimensions (BML model), for periodic and open boundaries. The traffic
flow model is given by a three-state CA on a square lattice. The CA model is
defined on a square lattice of $L\times L$ sites. Each site $\left(
i,j\right) $, with $1\leq i\leq L$ and $1\leq j\leq L$, contains either a
car moving upwards, a car moving to the right, or empty (Fig.1). In order
to take into account the anisotropic effect on BML\ model, we introduce in
this paper two parameters, $p_{ur}$ and $p_{ru}$, which exhibit,
respectively, the probability that an up-moving car change its direction to
become a right-moving car, and the probability that a right-moving car
change its direction to become an up-moving car. At the initial
configuration, cars are randomly distributed at the sites on the square
lattice in such manner that a uniform random number $\rho $ ($0\leq \rho
\leq 1$) is generated independently at each site and if $0\leq \rho \leq
\rho _{x}$ its site is occupied by a right car, if $\rho _{x}\leq \rho \leq
\rho _{x}+\rho _{y}$ its site is occupied by a upwards car and if otherwise
its site is empty.

In each time steps, the cars are randomly selected within sequential
dynamics, if the selected car is in right-moving (up-moving) state, it moves
to the right (up) unless the adjacent site on its right (upwards) hand side
is occupied by another car, which can be either an up or right driver. If it
is blocked by another car it does not move. After that, if the selected car
is in the right-moving (up-moving) state, its state is altered into
up-moving (right-moving) state with the probability $p_{ru}$ ($p_{ur}$).
Then, we perform computer simulations of the CA model starting with a
set of random initial conditions for the system size $L=10-500$, the
density $\rho =0.0-1.0$ of cars. Each run is obtained after $10000-50000$
time steps. After a transient period that depends on the system size, on the
random initial configuration and on the density of car, the system reaches
its asymptotic state. In order to compute the average of any parameter $u$ ($%
\left\langle u\right\rangle $), the values of $u\left( t\right) $ obtained
in the asymptotic state are averaged. \\
If we denote $\tau \left( i,j\right) $ the state of the site $\left(
i,j\right) $. The periodic boundaries case is defined by the conditions: 
\\
for  $1 \leq i\leq L: \tau \left( i,0\right) =\tau \left(i,L\right)$ and $ \tau \left( i,L+1\right) =\tau \left( i,1\right)$ \\
for  $1 \leq j\leq L: \tau \left( 0,j\right) =\tau \left(
L,j\right) $ and $\tau \left( L+1,j\right) =\tau \left( 1,j\right)$ \\
While for the open boundaries case, in each time steps, we visit the sites $%
\left( i,1\right) $ ($1\leq i\leq L$), each time that a site is empty then a
right-moving car is injected with the probability $\alpha $. While the
up-moving cars are injected in the sites $\left( 1,j\right) $ ($1\leq j\leq
L $) with the probability $\alpha $ when the site is empty. If an up-moving
(right-moving) car reaches one of the sites located in the upper (right) of
the lattice, i.e. the sites $\left( L,j\right) $ ( $\left( i,L\right) $ )
with $1\leq j\leq L$ ( $1\leq i\leq L$ ), it leaves the lattice with the
probability $\beta $.

In the following section we have used the parameters $v_{u}$, $v_{r}$ and $%
v_{g}$ defined , respectively, as the mean up-velocity, the mean
right-velocity and the mean velocity. In the periodic boundaries condition, $%
v_{u}$ ($v_{r}$) is the number of move performed by the up-moving
(right-moving) cars calculated in each time steps averaged over the
up-moving (right-moving) cars. The same procedure is carried out in order to
compute $v_{g}$ except that we average over all cars. In the open boundaries
condition, the parameters are calculated inside the square of length $2l+1$ (%
$l=6$) centered in the middle of the lattice (i.e. $\left( \frac{L+1}{2},%
\frac{L+1}{2}\right) $). We define the parameters $d_{u}$, $d_{r}$ and $d$ ($%
j_{u}$, $j_{r}$ and $j$), as the density (current) at the middle of the
lattice of , respectively, the up-moving cars, right-moving cars and all
types of cars. The currents $j_{u}$ and $j_{r}$ at the site $\left(
i,j\right) $ are defined, respectively, by $\left\langle u\left( i,j\right)
(1-g\left( i+1,j\right) \right\rangle _{l}$ and $\left\langle r\left(
i,j\right) (1-g\left( i,j+1\right) \right\rangle _{l}$, where $\left\langle
{}\right\rangle _{l}$ is the average over the square $\left( 2l+1\right)
^{2} $, while the global current is the summation of the both currents, $%
j=j_{u}+j_{r}$. The parameters $u\left( i,j\right) $, $r\left( i,j\right) $
and $g\left( i,j\right) $ are defined as the probability to find the site $%
\left( i,j\right) $ occupied, respectively, by the up-moving car,
right-moving car and any type of cars. Hereafter, we use the following
parameters, $\rho _{u}=\left\langle d_{u}\right\rangle $, $\rho
_{r}=\left\langle d_{r}\right\rangle $ and $\rho =\left\langle
d\right\rangle $.

\section{Simulations and results}

\subsection{Periodic boundaries}

We exhibit in Fig.2 the variation of the global mean velocity as a
function of the density for $p_{ur}=p_{ru}=0$, for different system sizes.
The system presents two different asymptotic state, which are separate by a
sharp dynamical transition. Before the transition, all cars move freely and
the average velocity is $\left\langle v_{g}\right\rangle =1$, while when the
transition occurs, they are all stuck and $\left\langle v_{g}\right\rangle =0$
separate rows of right and up cars along the diagonals from the upper-left
to the lower-right corners, this situation prevents the cars to move. As the
system size increases, the critical density $\rho _{c}$ tends to decrease
giving rise to sharper transition, and stabilize for high system sizes ($%
L\geq 300$).

In the isotropic case, i.e. $p=p_{ur}=p_{ru}$ ($p\neq 0$), Fig.3
presents the variation of $\left\langle v_{g}\right\rangle =\left\langle
v_{r}\right\rangle =\left\langle v_{u}\right\rangle $ as a function of the
density for different values of $p$. The sharp dynamical transition vanishes
and the mean velocity decreases monotonically with the density. In the case $%
p=1$, for low density ($\rho \leq 0.34$) the mean velocity is equal to $1
$, which means that each car moves without undergoing interactions with the
other cars, so the cars take routes that never overlapping. For $\rho
\geq 0.34$, the mean velocity decreases almost linearly with increasing
the density. For $p\neq 1$, the mean velocity decreases with increasing the
density, in this case the cars interact with each other since a car leaves
its oblique route and block other cars. By fixing the density, the mean
velocity increases with $p$. So, the growing of $p$ has for effect to avoid
the formation of the traffic jam since the cars leave the tails very fast,
this liberates the route for other cars which come from perpendicular
direction or moving behind it. In the anisotropic case (i.e., $p_{ur}\neq
p_{ru}$), we exhibit in the Figs.4a, 4b and 4c, respectively, the global,
up and right mean velocities versus the density for different values of $%
p_{ur}$ and taking $p_{ru}=1$ . The mean velocities (i.e., $\left\langle
v_{r}\right\rangle $, $\left\langle v_{u}\right\rangle $ and $\left\langle
v_{g}\right\rangle $) decrease with increasing the density, both the right
and up mean velocities vary with $p_{ur}$. Indeed $\left\langle
v_{r}\right\rangle $ increases with $p_{ur}$, while $\left\langle
v_{u}\right\rangle $ doesn't vary monotonically with varying $p_{ur}$.

In order to understand the variation of the velocities versus the anisotropy
for fixed density, one have to investigate the behavior of the mean
velocities versus $p_{ur}$ and $p_{ru}$. Since the symmetry of the system,
it suffice to study the variation $\left\langle v_{r}\right\rangle $ and $%
\left\langle v_{u}\right\rangle $ as a function of $p_{ur}$ instead of \
studying them versus both parameters. So, we exhibit the variation of $%
\left\langle v_{r}\right\rangle $ and $\left\langle v_{u}\right\rangle $ as
a function of $p_{ur}$ for different values of $p_{ru}$ in the Figs. 5 and
6 for, respectively, $\rho =0.3$ (low density) and $\rho =0.7$ (high
density). In the case $p_{ur}=0$, $\left\langle v_{u}\right\rangle =1$
(Fig.5a), there is only the up-moving cars in the lattice, and since the
system is in the moving phase, the up-moving cars move freely. This is not
the case at the jamming phase (Fig. 6a), in which the traffic jam tails of
up-moving cars form, then limiting the mean velocity of up-moving cars ($%
\left\langle v_{u}\right\rangle \neq 1$, for $p_{ur}=0$). By increasing $%
p_{ur}$ from $p_{ur}=0$, the mean up-velocity decreases at low values of $%
p_{ur}$, so, with the augmentation of $p_{ur}$ there is the formation of
right-moving cars in the lattice provoking the blockage of the up-moving
cars. The decrease is more important for low $p_{ru}$, in fact, for low
values of $p_{ru}$, the right-moving cars preserve their direction for long
time yielding the formation of more and more long traffic jam tails. At low
density, the up-velocity starts increasing from a value of $p_{ur}$ which
grows with $p_{ru}$. This augmentation of the up-moving velocity is the
consequence of the fact that the up-moving cars change their direction
frequently avoiding the grow of the local small tails in the upwards
direction. Since at low values of $p_{ru}$, the formation of the traffic jam
tails is more important, the augmentation of the up-moving velocity is
manifested from low values of $p_{ur}$. At high density, for $p_{ru}\leq
0.5$, the mean up-velocity preserve the same behavior noted at low density,
except the fall occurred at low $p_{ur}$ is more important for $p_{ru}=0.1$,
while for $p_{ru}>0.5$ the mean up-velocity decreases with increasing $%
p_{ur} $ for any value of $p_{ur}$. In fact, at high density, the number of
the empty sites is reduced, so the cars could not always leave the traffic
jam tails. This explain the fall of $\left\langle v_{u}\right\rangle $\ for
very low values of $p_{ur}$ for $p_{ru}=0.1$, indeed, by increasing $p_{ur}$
under these conditions the number of right-moving cars increases without
allowing to up-moving cars to leave the tails. For $p_{ru}\geq 0.5$, even
for high values of $p_{ur}$, $\left\langle v_{u}\right\rangle $\ decreases
with increasing $p_{ur}$. The up-moving cars in this case could not always
leave the tails when they change their state to become right-moving
vehicles, for high values of $p_{ru}$, they don't stay in the same state for
long time, they change their state to upward without leaving the tails. At
high values of $p_{ur}$, the variation of $\left\langle v_{u}\right\rangle $
is more and more weak with increasing $p_{ur}$ until becoming constant. In
fact, for high values of $p_{ur}$, even if $p_{ur}$ increases the situation
of the traffic jam does not change since there is the balance between the
formation of the obstacles (right-moving cars), and the rate of unblocking
the local clusters of traffic jam. The mean velocity of right-moving cars
increases with $p_{ur}$ for any value of $p_{ru}$, except,at low density, the decrease
occurred at lower $p_{ur}$ for $p_{ru}<0.5$ (Fig.5b) and at hight density when%
$\ p_{ru}\leq 0.5$ (Fig.6b). In fact, by increasing $p_{ur}$
the number of up-moving cars decreases, yielding the diminish of the
obstacles for the right-moving cars. For low values of $p_{ru}$, $%
\left\langle v_{r}\right\rangle $ diminish for very low values of $p_{ur}$
before undergoing an increase. In fact, under these conditions, as we have
quoted above, there is the formation of small number of right-moving cars,
which block the up-moving cars during their move, but $p_{ur}$\ is not so
higher to unblock the tails of the up-moving cars (the decrease of $%
\left\langle v_{u}\right\rangle $\ for low $p_{ur}$), this affects the
right-moving cars during their move which are slowed down by the tails of
up-moving cars. In the Fig. 7, we present the variation of the global,
right and up velocities as a function of $p_{ur}$ for $p_{ru}=0.5$ and $\rho
=0.7$. We note that for $p_{ur}<p_{ru}$ ($p_{ur}>p_{ru}$), the number of
up-moving (right-moving) cars is more important than the number of
right-moving (up-moving) cars, therefore the value of the mean global
velocity tends to the right (up) velocity. For $p_{ur}=p_{ru}$, the
velocities are equal.

\subsection{Open boundaries}

In this case, the system exhibits three phases; moving phase, jamming phase and maximal
current phase \cite{22}-\cite{24}. These phases are governed by three
factors; the flow of entering of cars, the flow of exiting the system and
the velocity of cars inside the network. In the case of the moving phase,
the current and the density don't depend upon the variation of $\beta $ with
taking $\alpha $ fixed, while the density and the current increase by
increasing $\alpha $ and taking $\beta $ fixed (Fig. 8). At the jamming
phase, the current and the density inside the lattice don't depend upon $%
\alpha $ for fixed $\beta $ (Fig.8), while by increasing $\beta $ for a
given $\alpha $, the current increases and the density decreases. At maximal
current phase, the current reaches its maximal value, and by varying $\alpha
$ and $\beta $, the flow remains unchanged. If the parameters of our model
(i.e., $p_{ur}$ and $p_{ru}$) favour the higher velocities (and
eventually high maximal flow), we need high values of $\alpha $ and $\beta $
in order to maintain the flow in its high level, which implies the shrink of
the maximal current phase zone. In the isotropic case ($p=p_{ur}=p_{ru}$),
the velocities increase with $p$, so the maximal current phase zone is
contracted to high values of $\alpha $ and $\beta $ (Fig. 9). At the moving
phase, the density $\rho $ does not depend upon $p$ for low values of $%
\alpha $, while at higher $\alpha $, at vicinity of the transition, the
density increases with decreasing $p$ (Fig. 8). In fact, for a weak values
of $p$, there is the formation of small traffic jam clusters which block the
cars inside the lattice, for very low density (i.e. very low $\alpha $),
such clusters don't form and the cars move freely even at very low values of
$p$. Like in the asymmetric exclusion model (ASEP) \cite{22}-\cite{24}, at
the moving phase close to the transition to jamming phase, the system is
divided into two regions, namely, the higher density region at the exit of
the system, in our model this region is located at vicinity of the
upper-right exit (Fig. 10), and the lower density region elsewhere. For low
values of $p$, the system is blocked rapidly with the formation of local
traffic jam clusters, hence the density inside the higher density region
decreases. By increasing $p$, the traffic jam clusters are unblocked on
filling the empty sites located between these clusters, which increases the
density inside the higher density region. Therefore, the area (density) of
the higher density region increases (decreases) with decreasing $p$. Like in
the ASEP model, the transition between the moving and jamming phases occurs
when the higher density region invade the lattice. Since at high values of $%
p $, the higher density region is more compact, and the lower density is
more empty, the cars arrive more rapidly to the higher density region.
Increasing $\alpha $ near the transition, the higher density region
propagate inside the lattice and occupy the remaining exits, by increasing $p
$ above $p=0.5$, there is the equilibrium between the large surface of the
higher density region at low $p$ ($p\geq 0.5$) caused by the formation
of clusters, and the great velocity of cars moving forward this region at
high $p$, which accelerate its propagation, so the transition point $\alpha
_{c}$ does not vary with the variation of $p$ above $p=0.5$. As we have
quoted above, the density of the higher density region increases with $p$,
and since this region occupy the lattice in the jamming phase , the density
increases with $p$ in this phase. The transition between the high density
and the moving phases is a first order transition, while the transition
between the moving and jamming phases and the maximal current phase are a
second order transition, except in the case $p=1$, for which the transitions
are a first order transition. The first order transition line between the
jamming and moving phases exhibits a curvature which is due to the dimension
effect. Such result is obtained in the NaSch model in one-dimension with
open boundaries but in the case $v_{\max }>1$ \cite{20}. In fact, in the
both cases the car has more possibilities of move, in the NaSch model it has
$v_{\max }$ opportunities of move, while in our case the car could move
either right or upwards. In contrast with the transition from moving to
jamming phases, the transition between the moving phase and the maximal
current phase arises without the formation of the higher density region in
the exit of the system. The density inside the lattice increases
monotonically, except that the rate of the augmentation increases with
moving from the entrance to the exit. In the case $p=1$, with the absence of
the clusters, the augmentation of the density is not important at vicinity
of the transition, and since the density in the maximal current phase is
larger because of the absence of the great empty spaces, the intersection
between these two situations at the transition reflect an unstable
equilibrium state which indicate a first order transition. The first order
transition occurs at $\alpha <\beta $, indeed, the system reaches its
jamming state at low densities because of the mutual blockage of cars moving
in different directions.

In order to study the anisotropic case, i.e. $p_{ur}\neq p_{ru}$, we exhibit
in the Figs. 11 and 12 the variation of $\rho $ as a function of $\alpha $
and $\beta =0.4$ for, respectively, $p_{ru}=0.1$ and $p_{ru}=1$ and various
values of $p_{ur}.$ In the case $p_{ur}>p_{ru}$ ($p_{ur}<p_{ru}$), the
up-moving (right-moving) cars entering on the bottom (left) of the lattice
change their direction at vicinity of that entry to become the right-moving
(up-moving) cars, more $p_{ur}$ ($p_{ru}$) is larger than $p_{ru}$ ($p_{ur}$%
) more the move of cars is carried out at vicinity of bottom (left) entry in
the right (up) direction. This augmentation of the density in the entrance
(Fig. 13) prevents other cars to enter via this side, which implies the
diminish of the density inside the lattice at the moving phase. This
behavior in the entrance caused by the difference between $p_{ur}$ and $%
p_{ru}$ delay the fill of the system, so as the difference is important as
the transition occurs at high $\alpha _{c}$. At the jamming phase, the
difference between the rates $p_{ur}$ and $p_{ru}$, cause an abundance of
one type of cars in relation to other type, which limit the blockage between
different type of cars with decreasing the empty spaces, hence the
augmentation of the density. At higher values of $p_{ru}$ (i.e., $p_{ru}=1$)
and varying $p_{ur}$, the system vary between the situation in which there
is particularly the up-moving cars inside the lattice, and the situation in
which there is as many up-moving cars as right-moving ones without the
formation of traffic jam tails, the density between these both situations
does not vary considerably. While in the second case (i.e., $p_{ru}=0.1$)
and varying $p_{ur}$, the system vary between the situation in which there
is especially the right-moving cars inside the lattice, and the situation in
which there is as many right-moving cars as up-moving ones with the
formation of traffic jam tails yielding the diminish of the density provoked
by the spaces between the clusters. Indeed, the variation of the density
undergone at high density with the variation of $p_{ur}$ is more important
in the second case (i.e., $p_{ru}=0.1$). The zone of the maximal current
phase shrink with increasing $\left| p_{ru}-p_{ur}\right| $ (Figs. 14 and
15). In fact, as we have quoted above, having a greater difference between $%
p_{ru}$ and $p_{ur}$ means the formation of a condensate band in one
entering side which prevent the cars to enter the lattice. So, we have to
increase $\alpha $ to overcome the traffic jam in the entrance, in order to
obtain the density corresponding to the maximal current, hence the
transition line between the moving phase and the maximal current phase
increases. And as the density increases with the augmentation of $\left|
p_{ru}-p_{ur}\right| $ at the jamming phase, one must increase $\beta $ in
order to drain the higher number of cars, which provoke the augmentation of
the transition line between the jamming phase and the maximal current phase.
This variation of the transition line, with varying $\beta $ and fixed value
of $\alpha $, is not important in the case $p_{ru}=1$ since the density does
not vary considerably at high density by varying $p_{ur}$. The density
profile in the middle line (i.e. $i=L/2$ ( $j=L/2$) and varying $j$ ($i$) )
or in the oblique line from the left-bottom corner to right-upper corner,
decays in the maximal current phase with an exponent $\gamma \approx 0.20$
(Fig. 16), both in the anisotropic and isotropic cases. This means that the
model belong to another universality class than the models studied in one
dimension, such as the ASEP ($\gamma =\frac{1}{2}$) \cite{22},\cite{23} and
the NaSch model for $v_{\max }>1$ ($\gamma \approx \frac{2}{3}$) \cite{20}.

\section{Conclusion}

In this paper, we have established the anisotropy effect of directions of
move on the cellular automaton of two-dimensional traffic flow (i.e. BML
model), in the periodic and open boundaries conditions. We have shown in the
periodic boundaries conditions that the first order jamming transition
disappears when the cars could change their direction of move every time
steps. The cars of different type play the role of obstacles for each other,
by increasing the probability to change the direction from right (upward) to
upward (right), $p_{ru}$ ($p_{ur}$) by taking $p_{ur}$ ($p_{ru}$) fixed, the
mean velocity of up-moving (right-moving) cars increases, except the slight
decrease at low $p_{ru}$ ($p_{ur}$), while the mean velocity of right-moving
(up-moving) cars decreases before undergoing an augmentation. In the open
boundaries conditions, the system exhibits three phases, namely, moving
phase, jamming phase and maximal current phase. The first order transition
between the moving phase and the jamming one occurs at $\alpha <\beta $. The
passage from first to second order transition occurs by decreasing the
anisotropy. The zone of the high density (low density) phase contracts
(expands) by the augmentation of the anisotropy. When the car change its
direction of move every time steps, in the isotropic case, the transition
from the jamming phase (or moving phase) to the maximal current phase is a
first order transition instead of the second order one. The density profile
in the middle line decays in the maximal current phase with an exponent $%
\gamma \approx \frac{1}{4}$, both in the anisotropic and isotropic cases.

\newpage \textit{Figure captions:}

Fig.1 : Schematic illustration of the traffic flow on the square lattice.
There are two types of drivers: the right-moving car going to the right and
the up-moving car going to the upwards. The right cars are presented by the
symbol ($\Box $) and the upwards cars are presented by the symbol ($%
\triangle $), the arrows indicate the cars which could move.

Fig.2: The variation of the global mean velocity $\left\langle
v_{g}\right\rangle $ versus the density for different values of the lattice
size $L$.

Fig. 3: The variation of the global mean velocity $\left\langle
v_{g}\right\rangle $ versus the density for different values of $%
p=p_{ur}=p_{ru}$ ($L=100$).

Fig. 4: The variation of a) the global mean velocity $\left\langle
v_{g}\right\rangle $, b) the up mean velocity $\left\langle
v_{u}\right\rangle $, c) the right mean velocity $\left\langle
v_{r}\right\rangle $, versus the density for different values of $p_{ur}$,
and $p_{ru}=1$ ($L=100$).

Fig. 5: The variation of a) the up mean velocity $\left\langle
v_{u}\right\rangle $, b) the right mean velocity $\left\langle
v_{r}\right\rangle $, versus $p_{ur}$ for different values of $p_{ru}$, and $%
\rho =0.3$ ($L=100$).

Fig. 6: The variation of a) the up mean velocity $\left\langle
v_{u}\right\rangle $, b) the right mean velocity $\left\langle
v_{r}\right\rangle $, versus $p_{ur}$ for different values of $p_{ru}$, and $%
\rho =0.7$ ($L=100$).

Fig. 7: The variation of the mean velocities $\left\langle
v_{g}\right\rangle $, $\left\langle v_{u}\right\rangle $ and $\left\langle
v_{r}\right\rangle $ versus $p_{ur}$ for $p_{ru}=0.5$ and $\rho =0.7$ ($%
L=100 $).

Fig. 8: The variation of a) $\rho $ and b) $J$, as a function of $\alpha $
for different values of $p$ for $\beta =0.4$ ($L=101$).

Fig. 9: phase diagram $\left( \alpha ,\beta \right) $ for different values
of $p$, the continuous (dashed) line presents the first order (second order)
transition .

Fig. 10: Schematic configuration of the system in the isotropic case for $%
\beta =0.4$ and $\alpha =0.148$ and a) $p=0.5$, b) $p=1$. The up-moving cars
are indicated by the vertical bar and the right-moving cars by the
horizontal bar ($L=60$).

Fig. 11: The variation of $\rho $ as a function of $\alpha $ for different
values of $p_{ur}$ with $\beta =0.4$ and $p_{ru}=0.1$ ($L=101$).

Fig. 12: The variation of $\rho $ as a function of $\alpha $ for different
values of $p_{ur}$ with $\beta =0.4$ and $p_{ru}=1$ ($L=101$).

Fig. 13: Schematic configuration of the system in the anisotropic case for $%
\beta =0.4$, $\alpha =0.1$, $p_{ru}=0.1$ and $p_{ur}=0.7$. The up-moving
cars are indicated by the vertical bar and the right-moving cars by the
horizontal bar ($L=60$).

Fig. 14: phase diagram $\left( \alpha ,\beta \right) $ for different values
of $p_{ur}$, $p_{ru}=0.1$, the continuous (dashed) line presents the first
order (second order) transition .

Fig. 15: phase diagram $\left( \alpha ,\beta \right) $ for different values
of $p_{ur}$, $p_{ru}=1$, the continuous (dashed) line presents the first
order (second order) transition .

Fig. 16: The variation of the density profile, (a) $\rho \left( \frac{L}{2}%
,i\right) $ versus $i$ (the horizontal middle line) and (b) $\rho \left(
i,i\right) $ versus $i$ (the oblique line) at the maximal current phase for $%
L=500$.

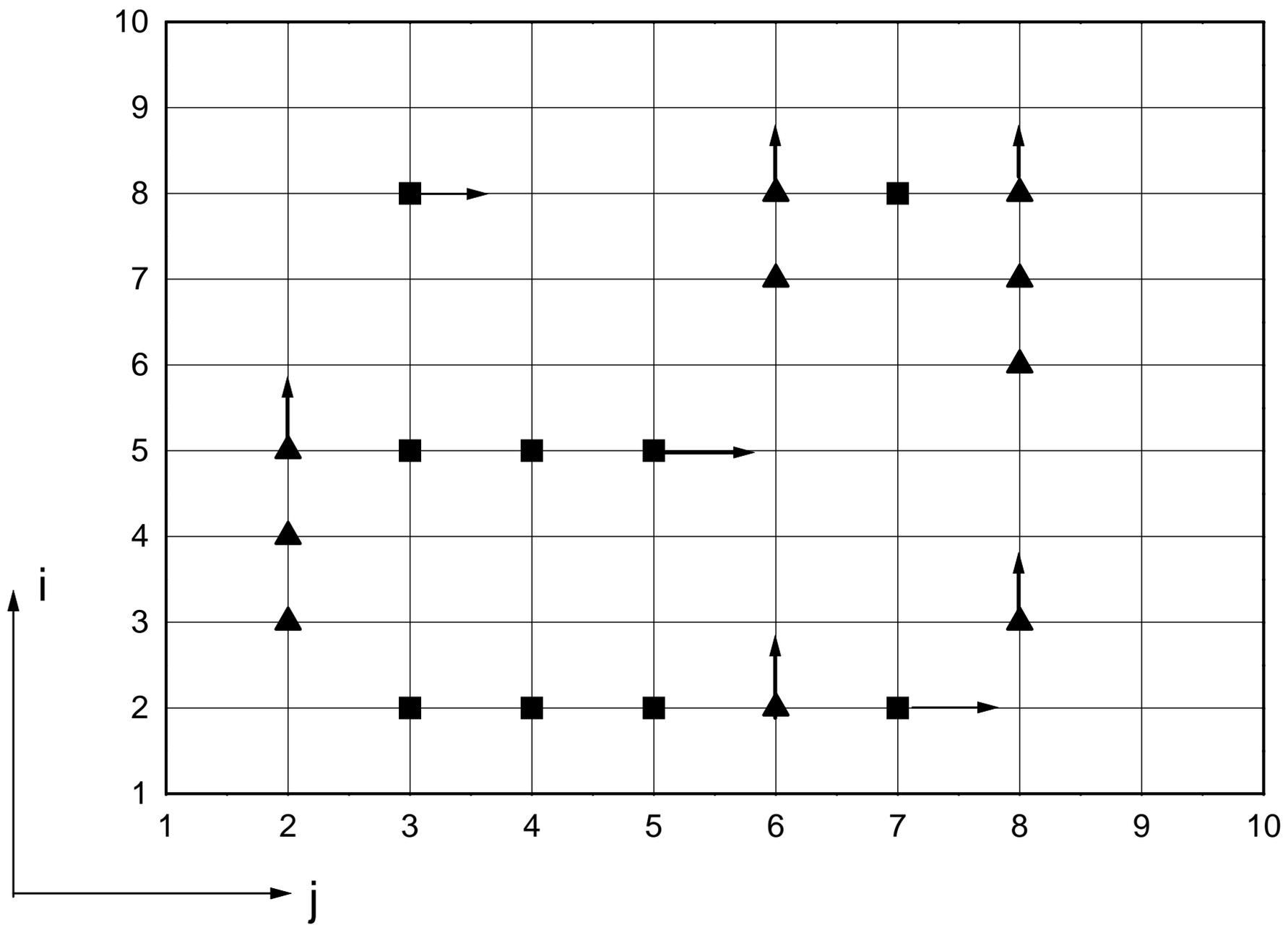
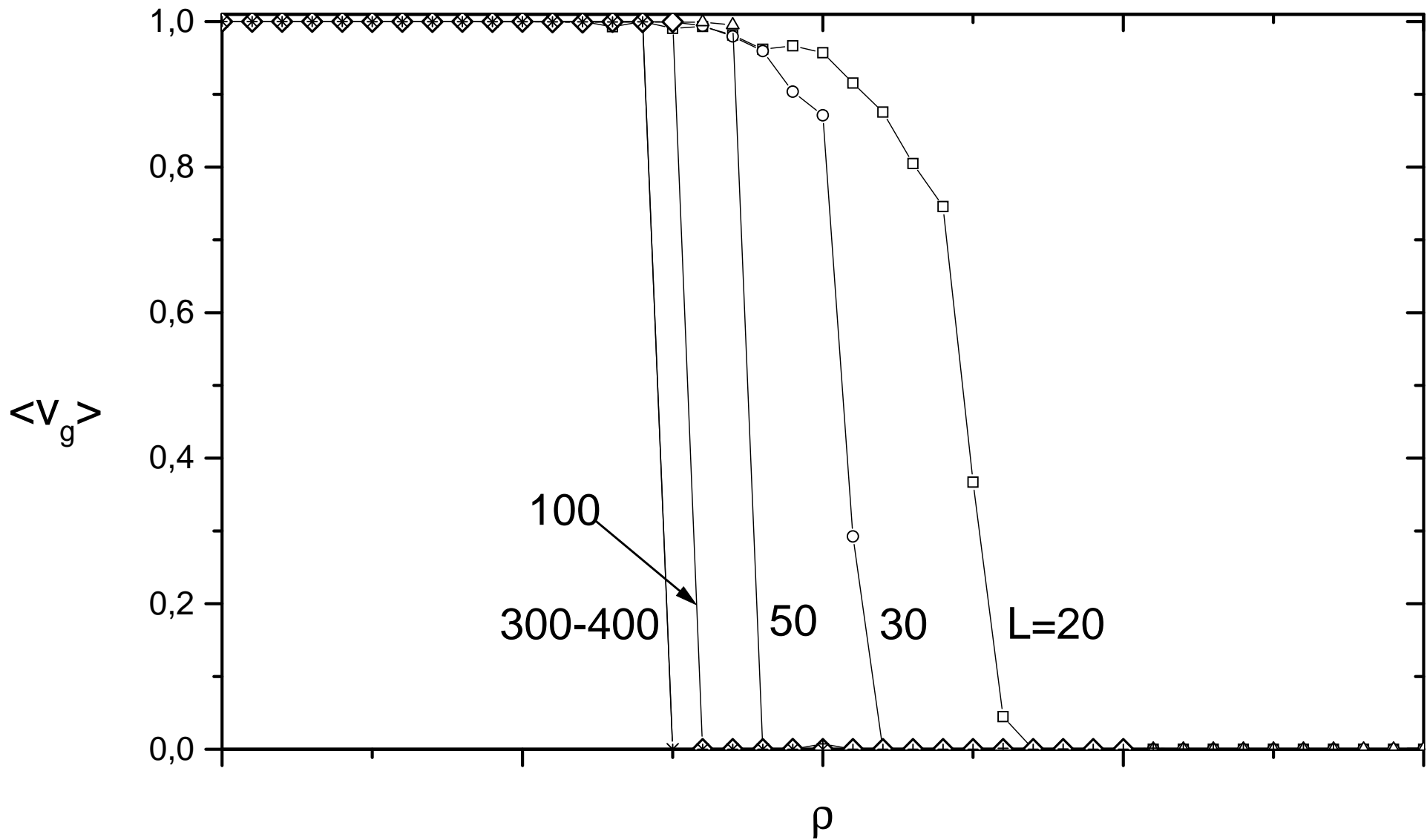
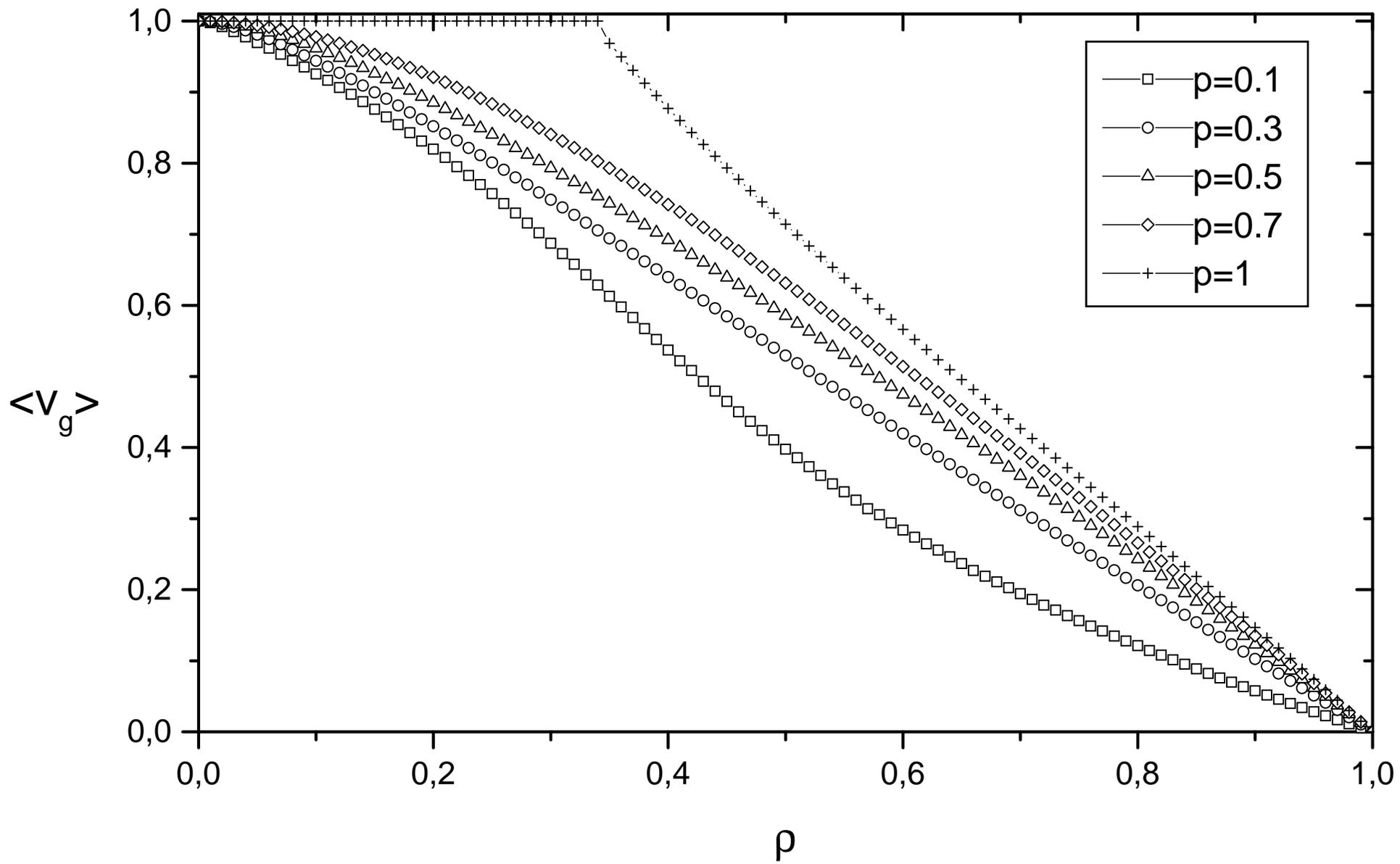
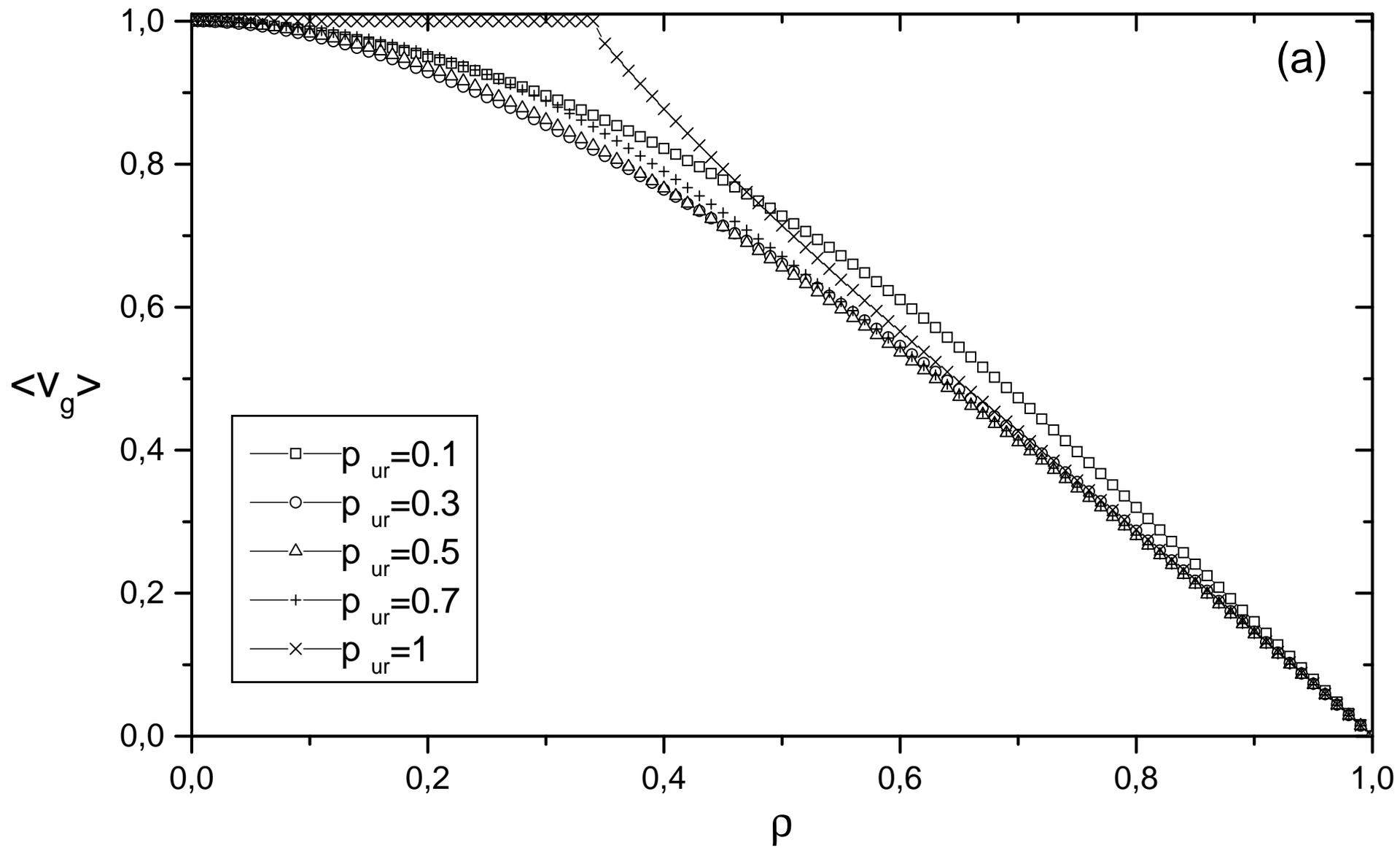
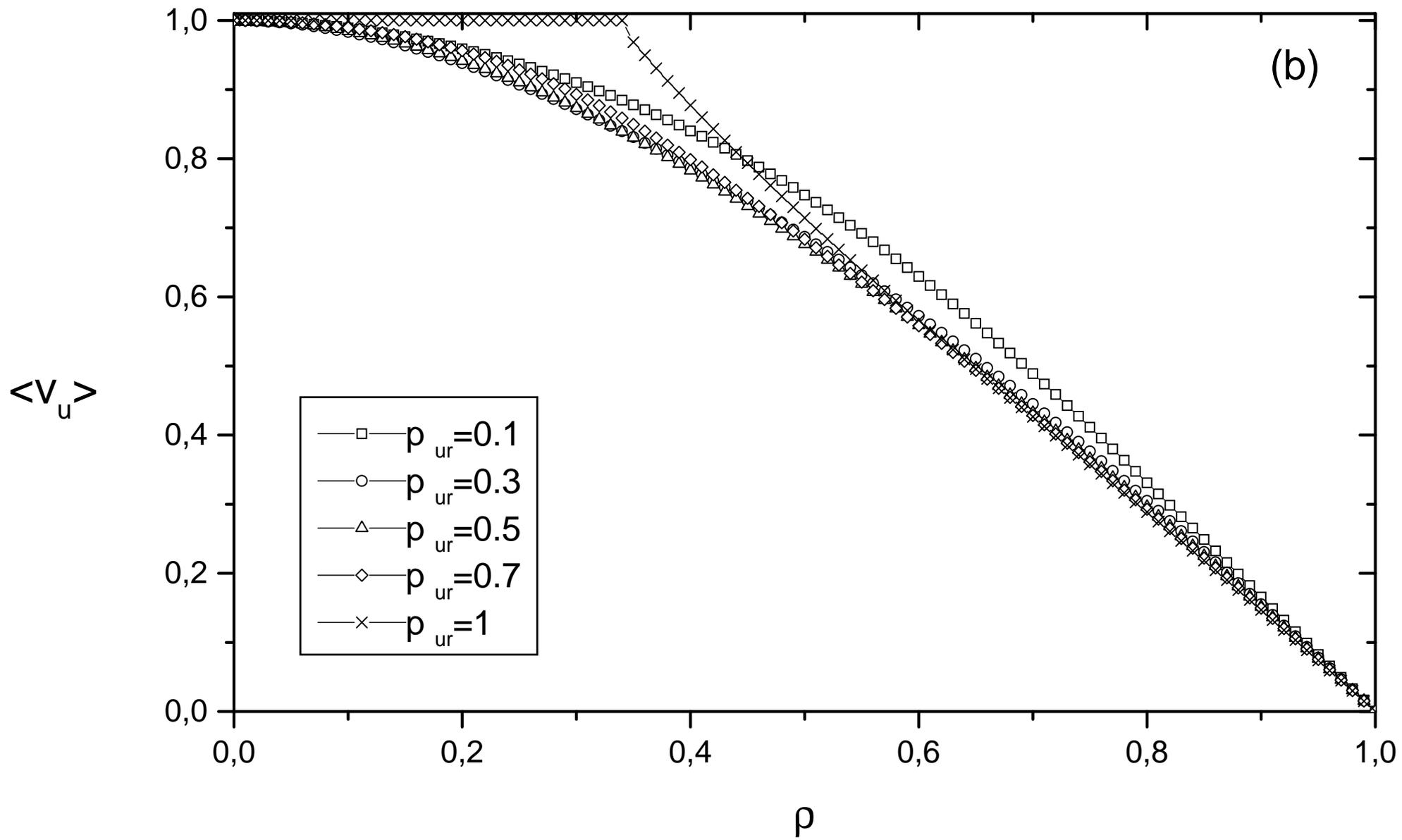
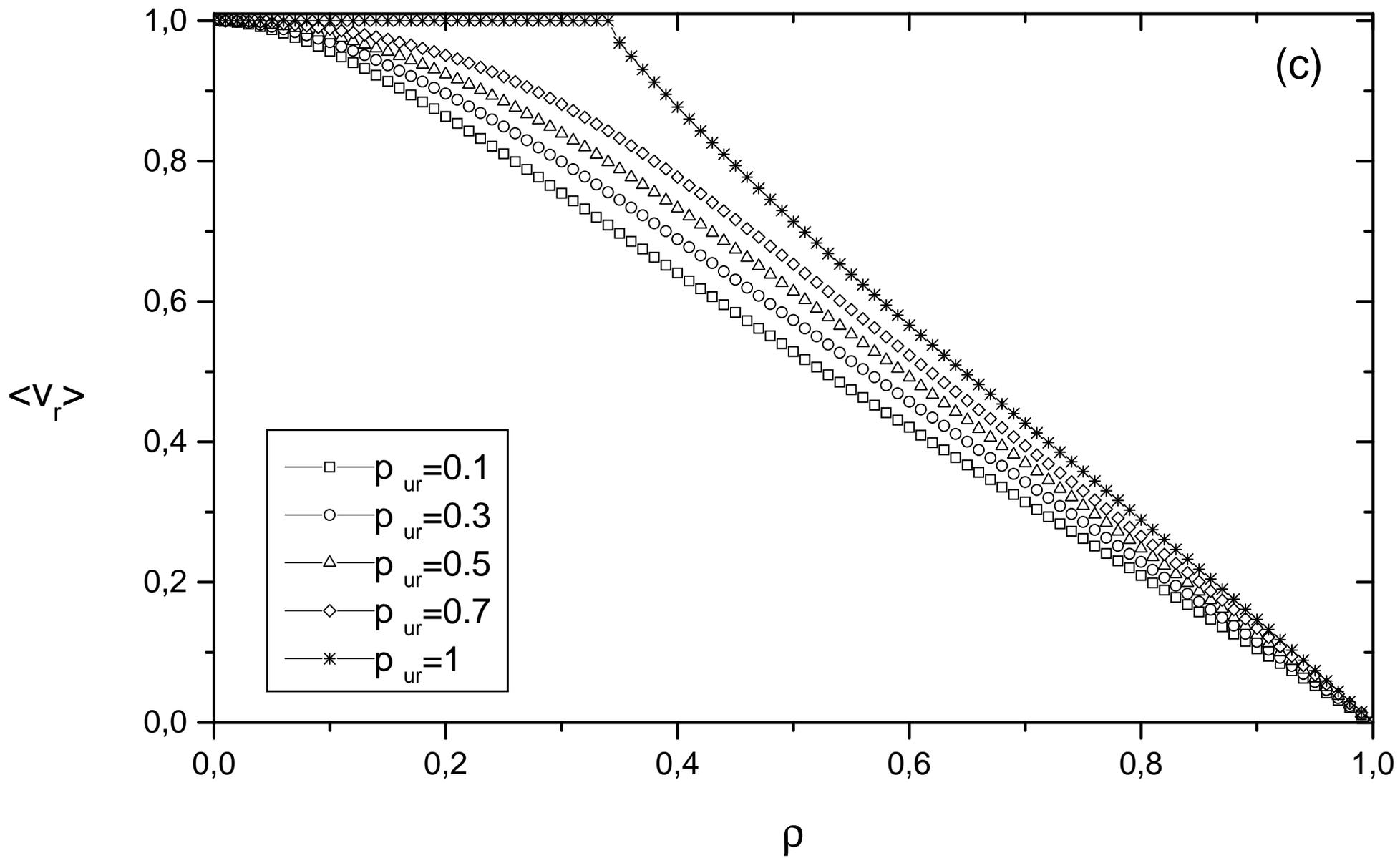
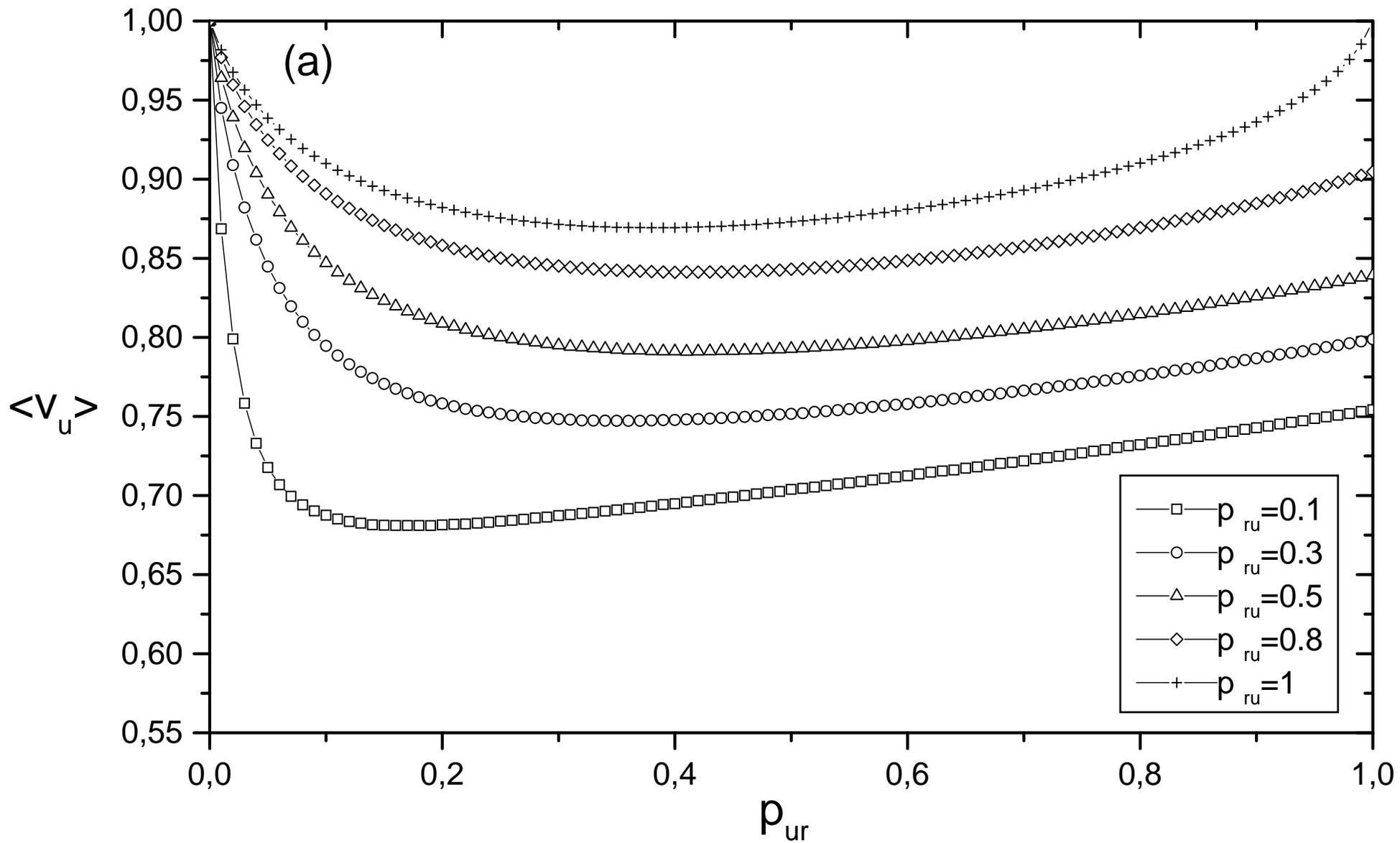
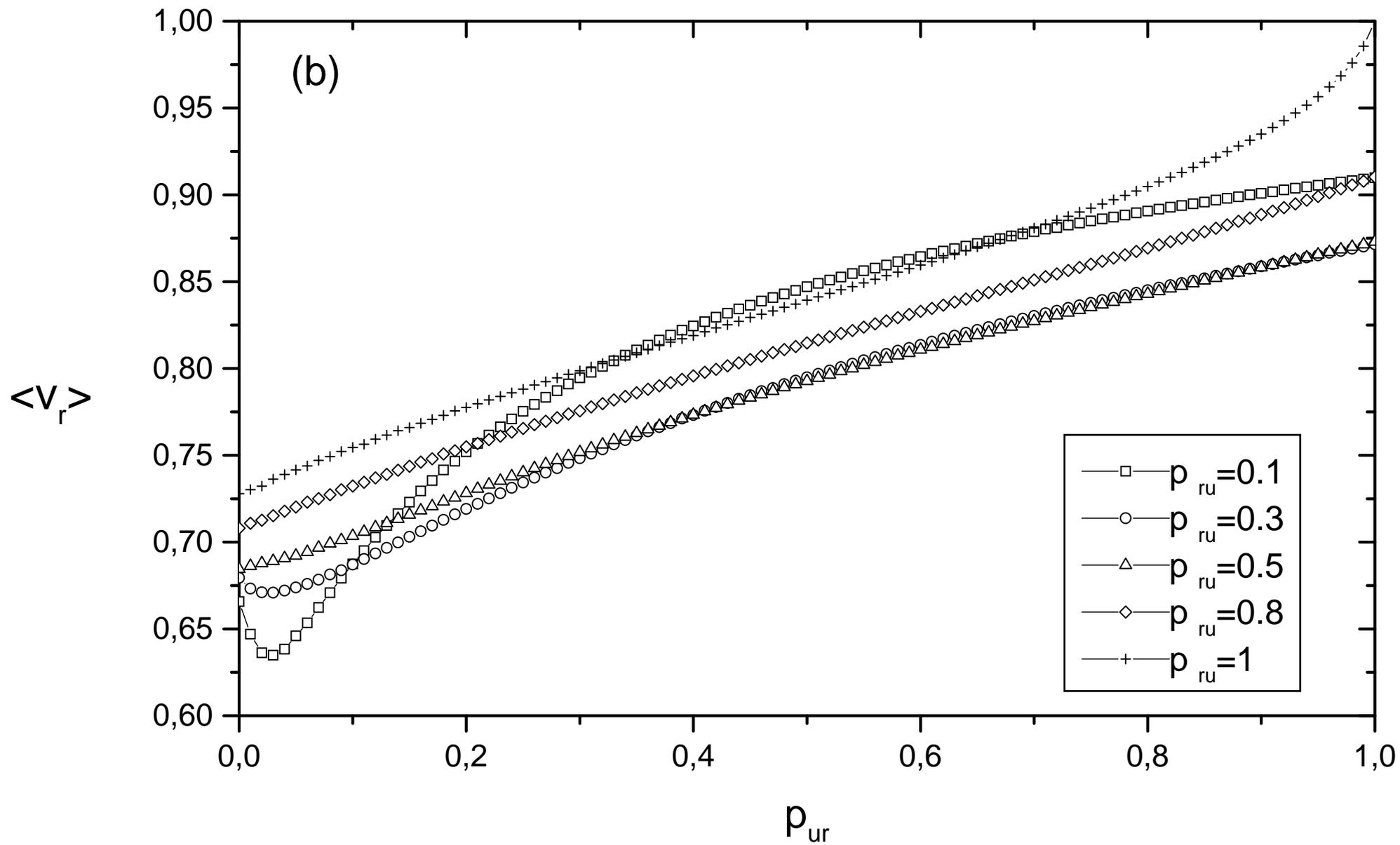
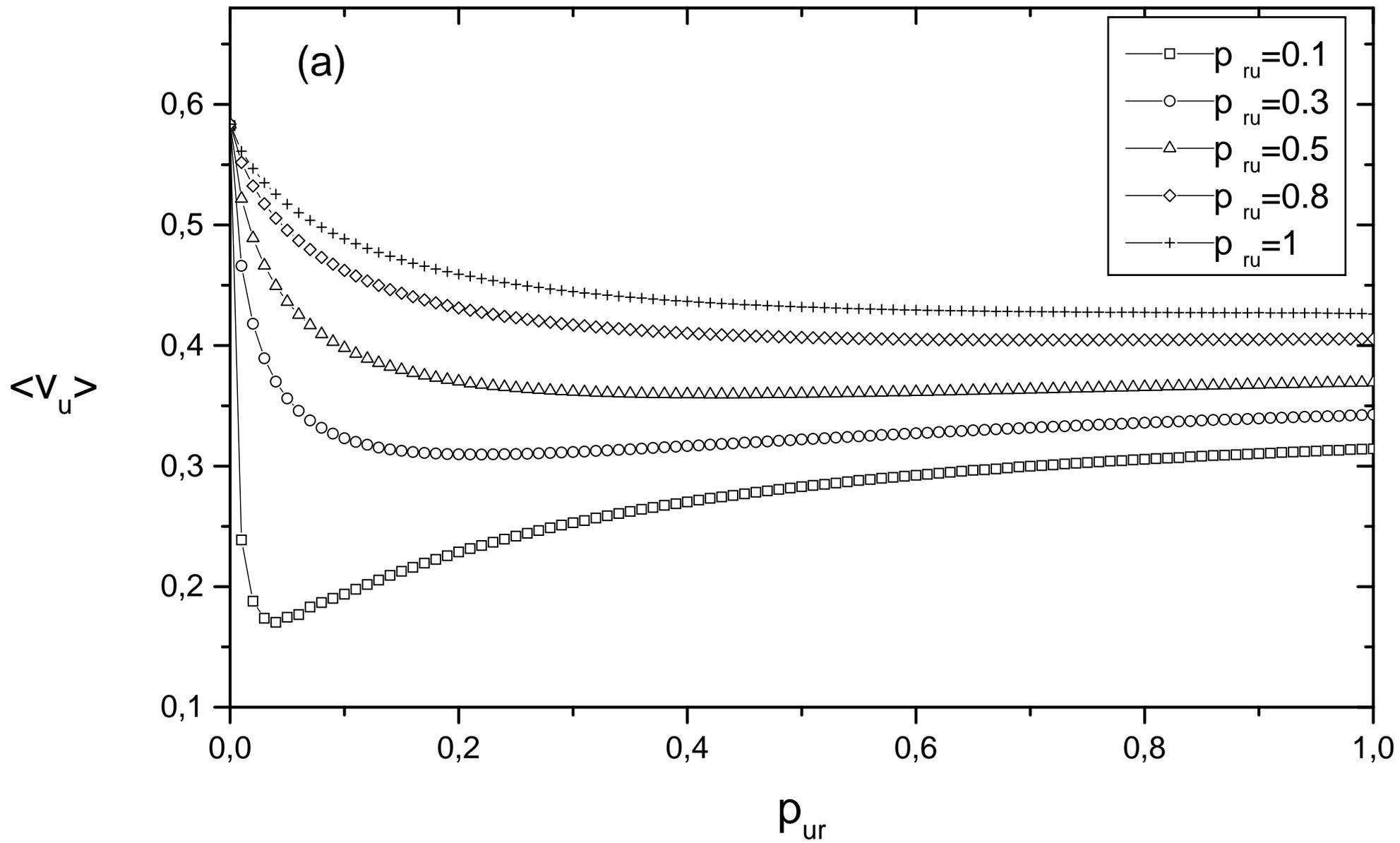
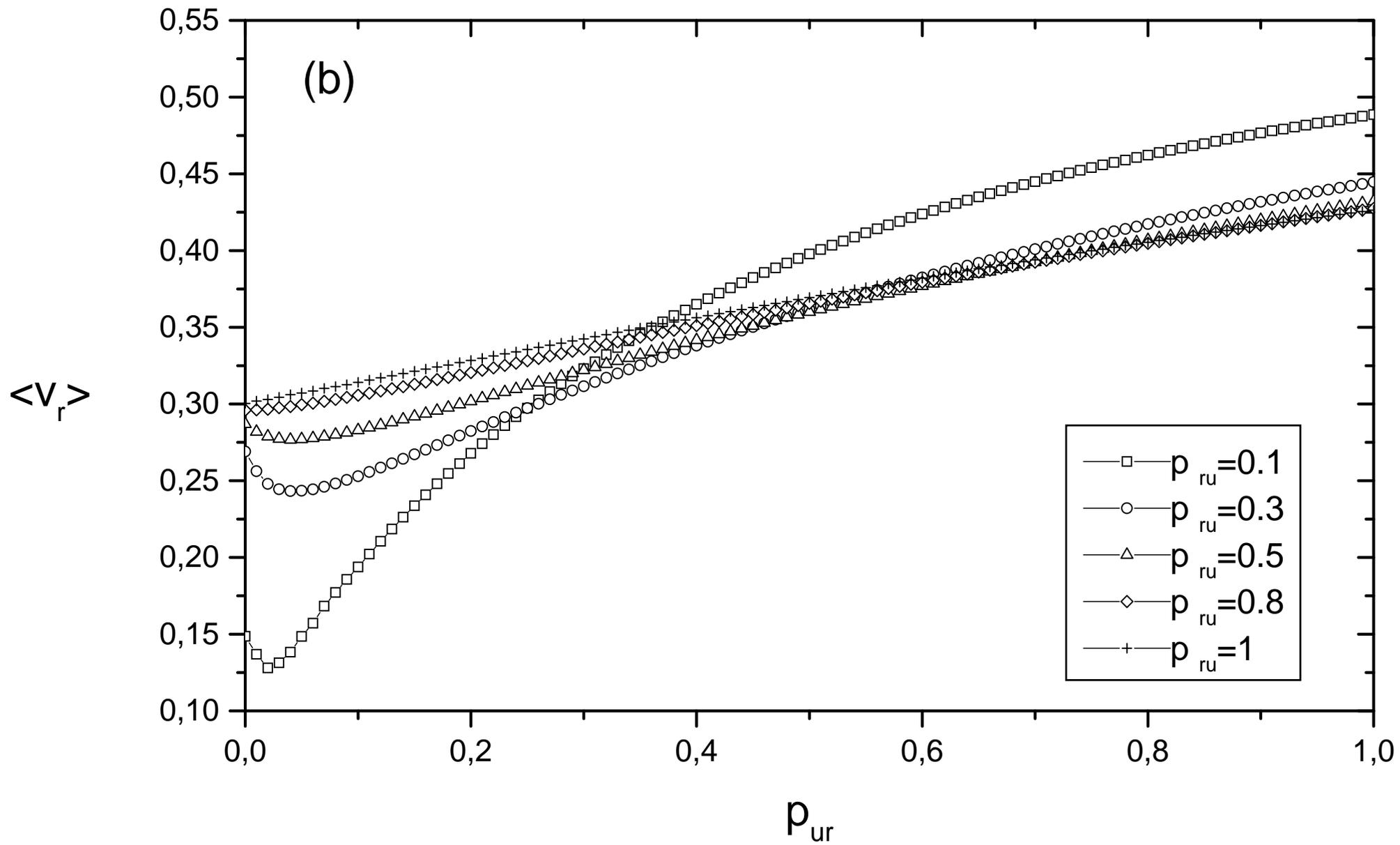
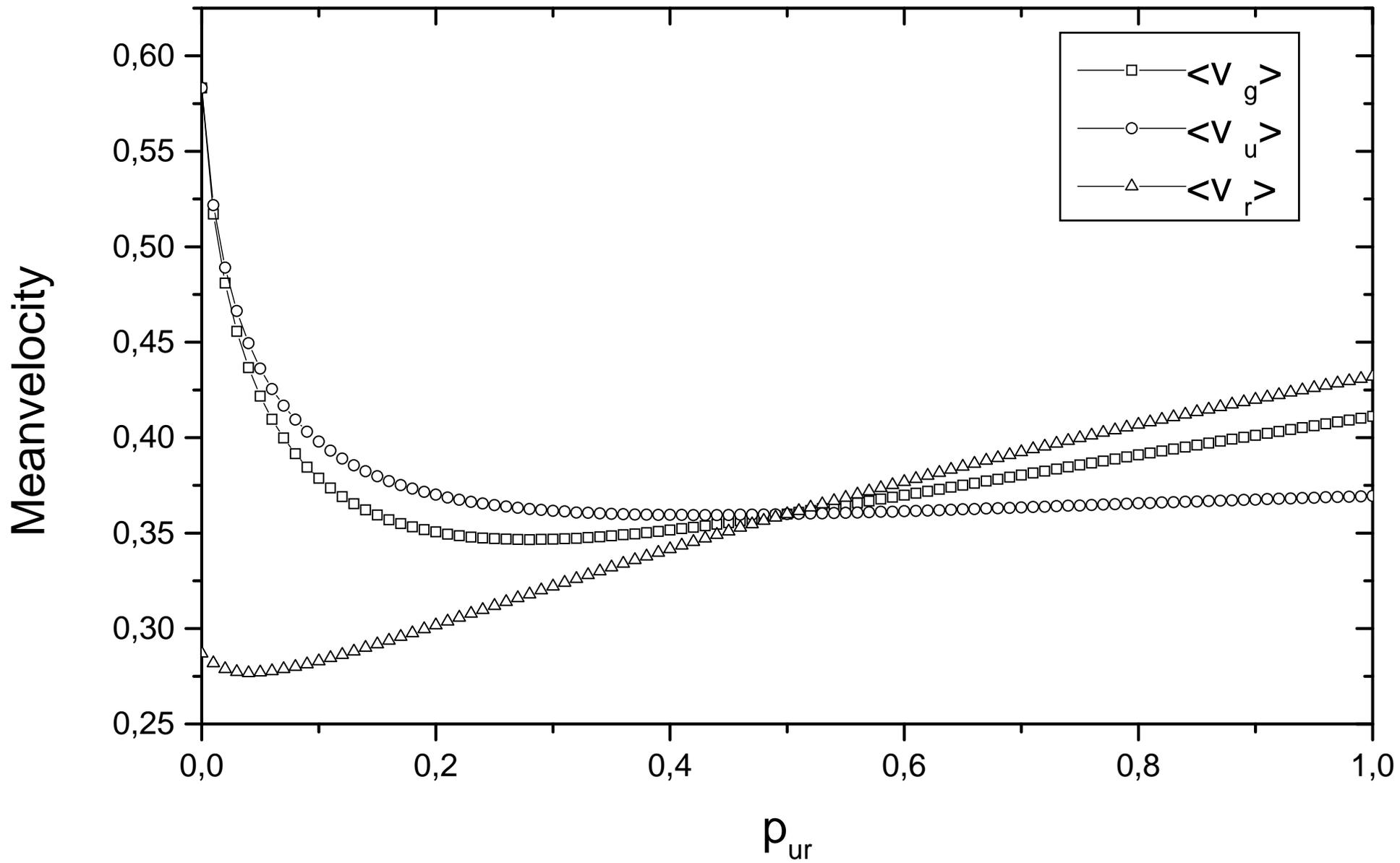
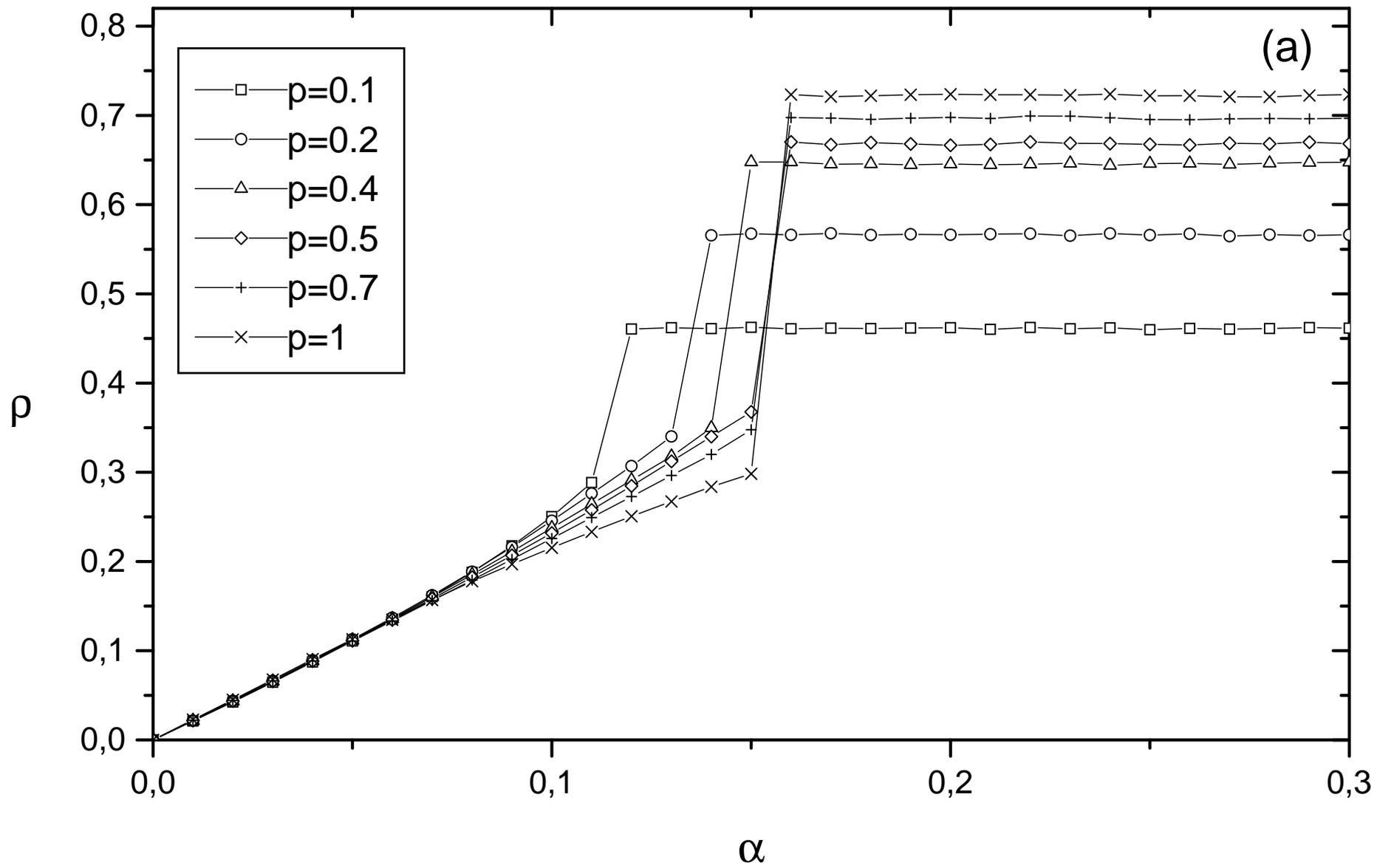
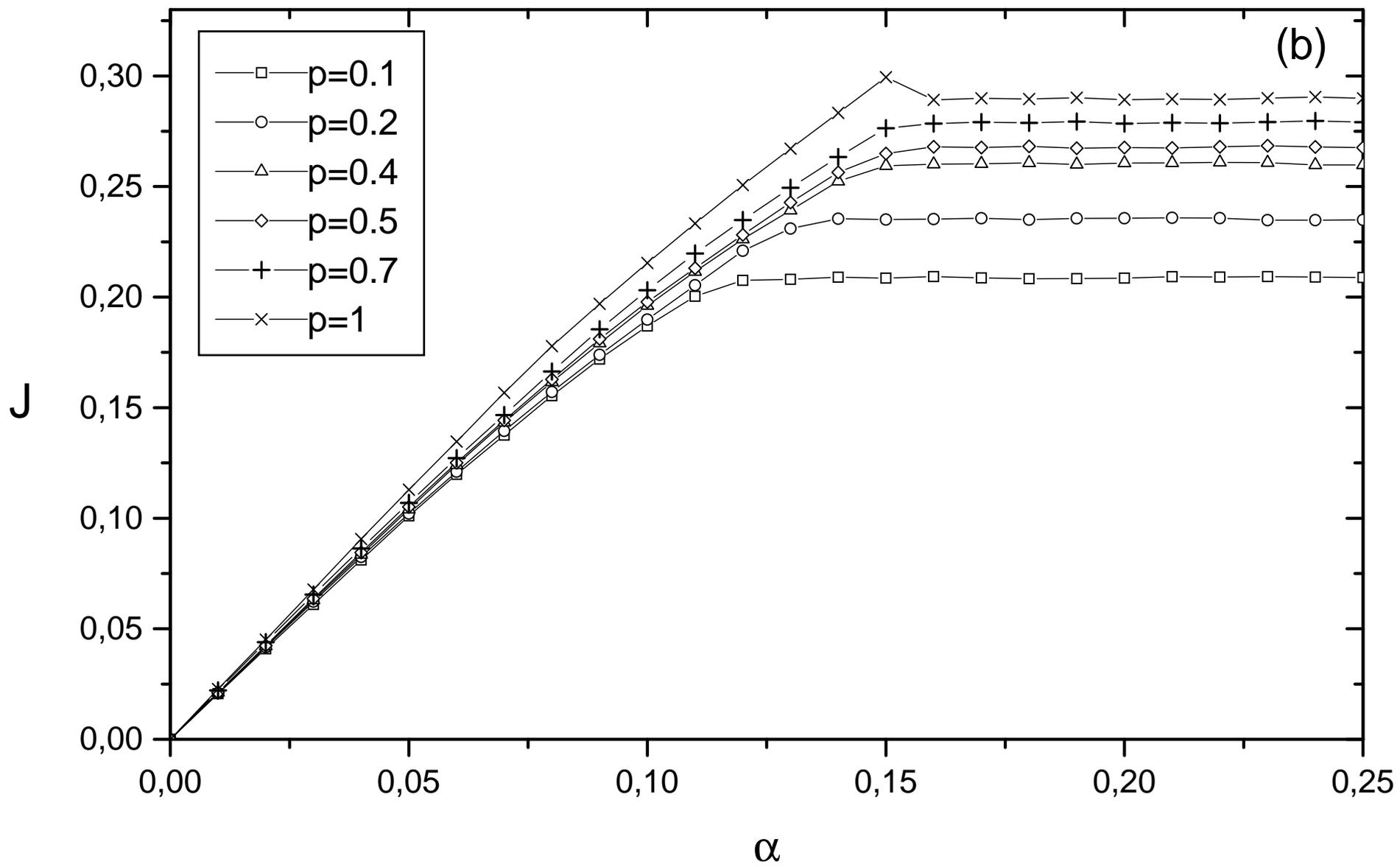
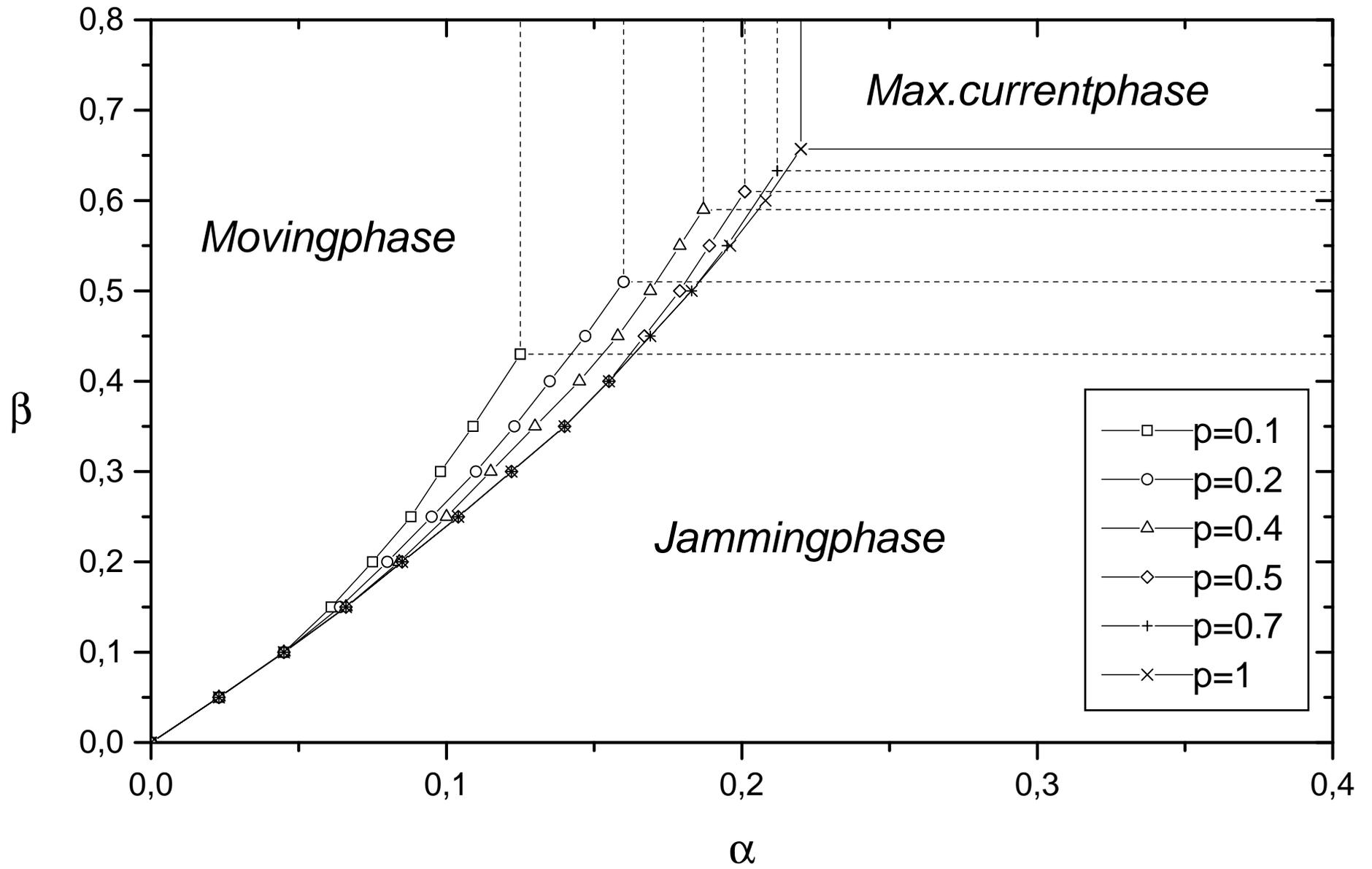
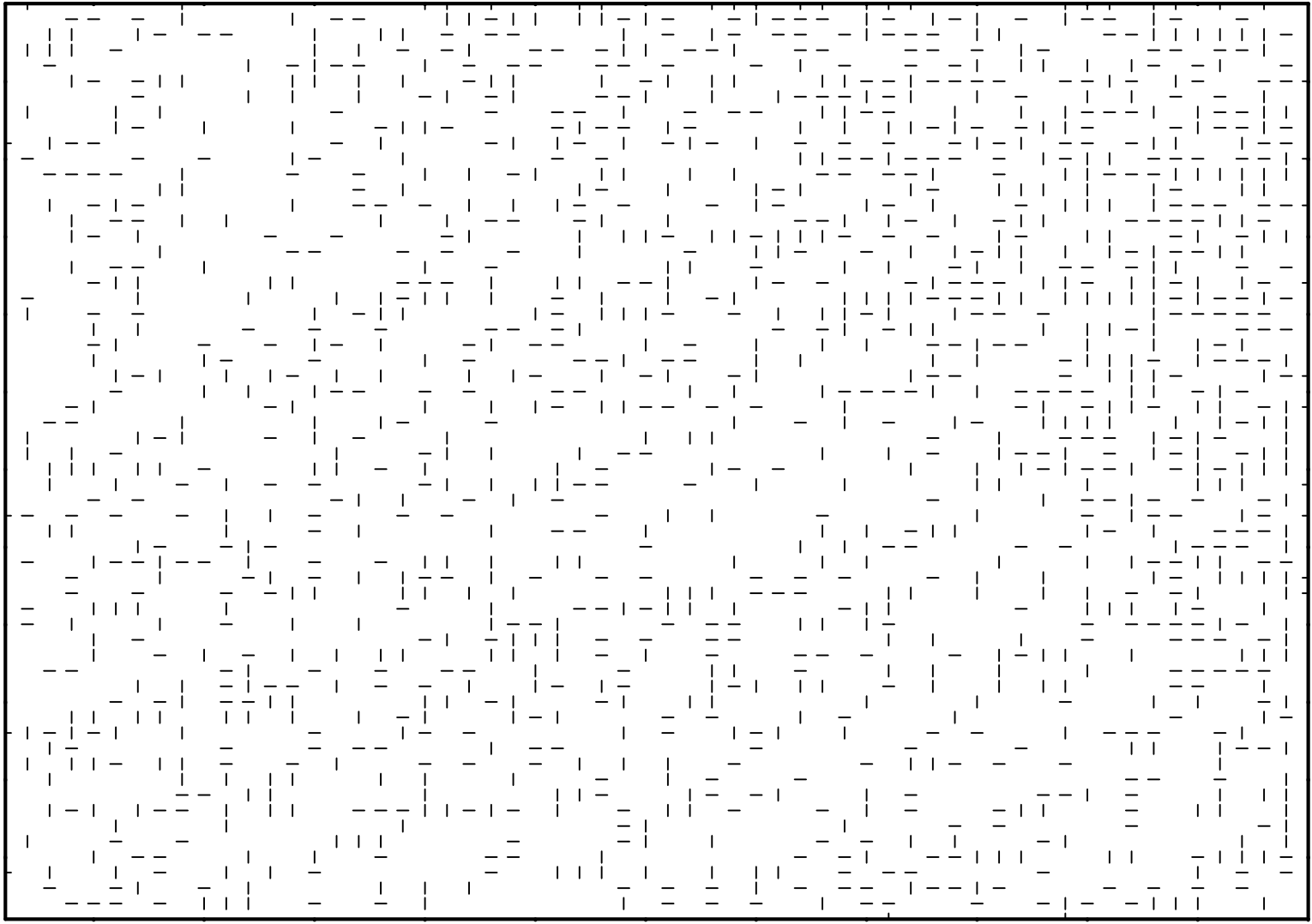
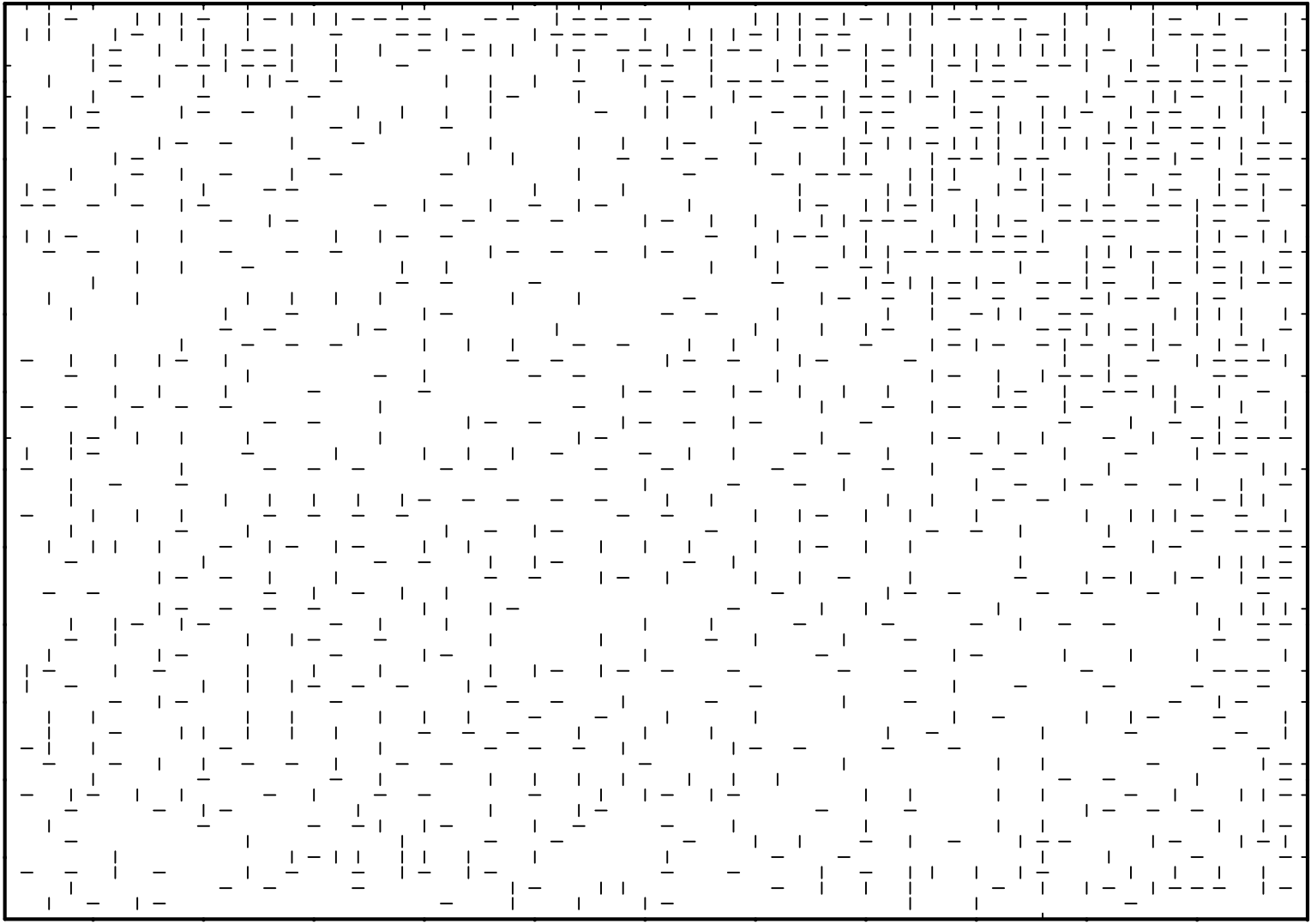
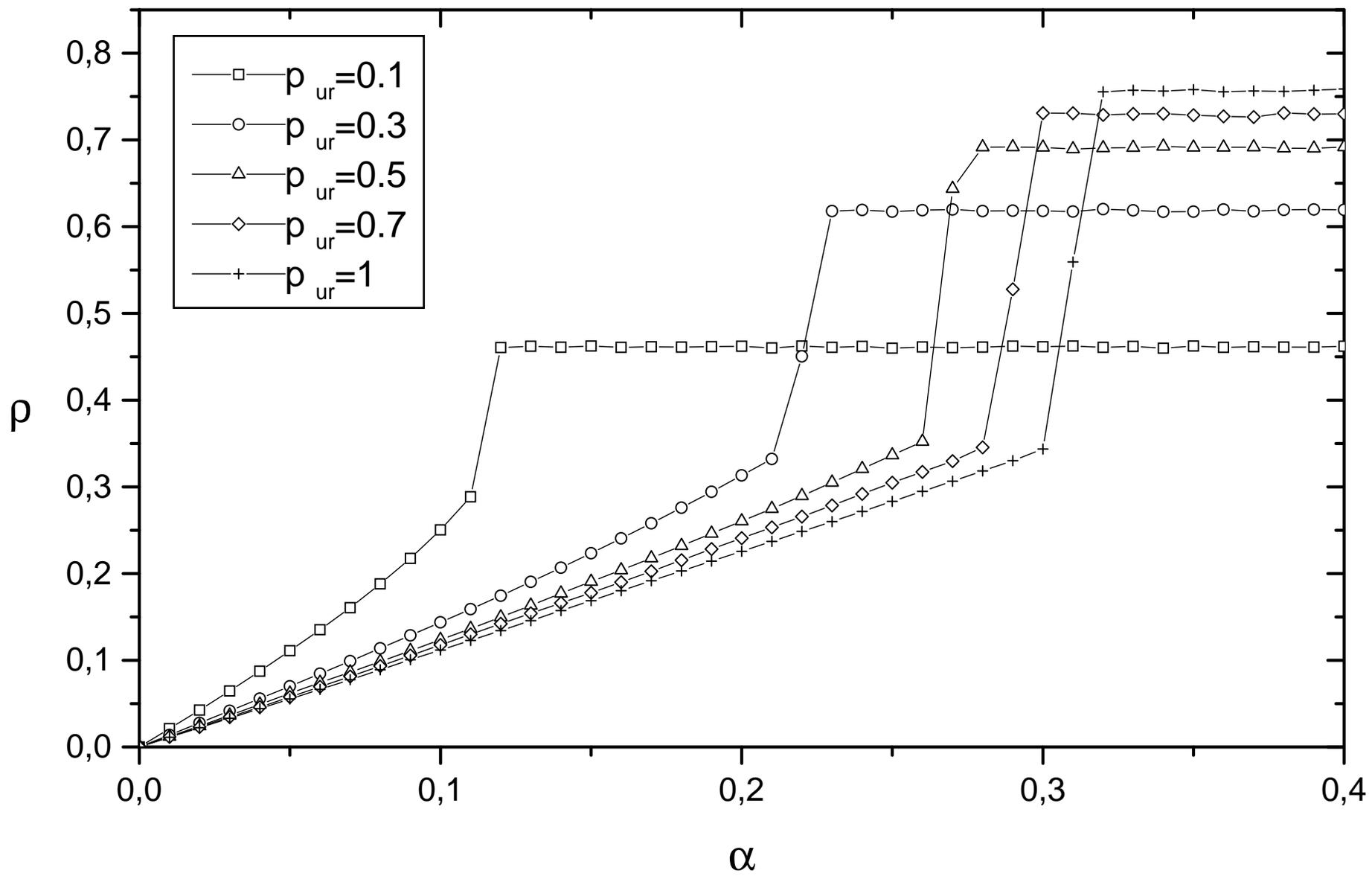
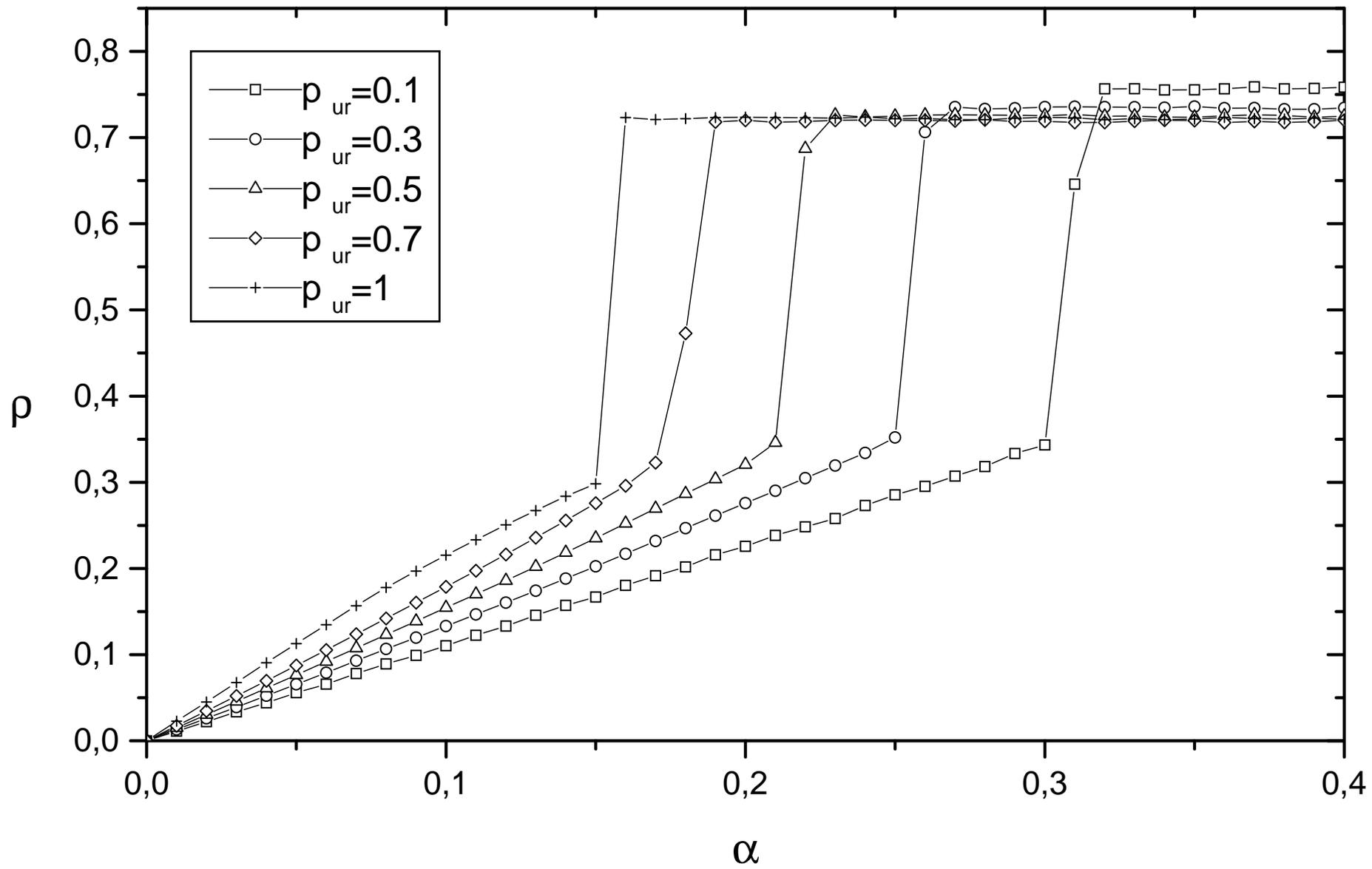
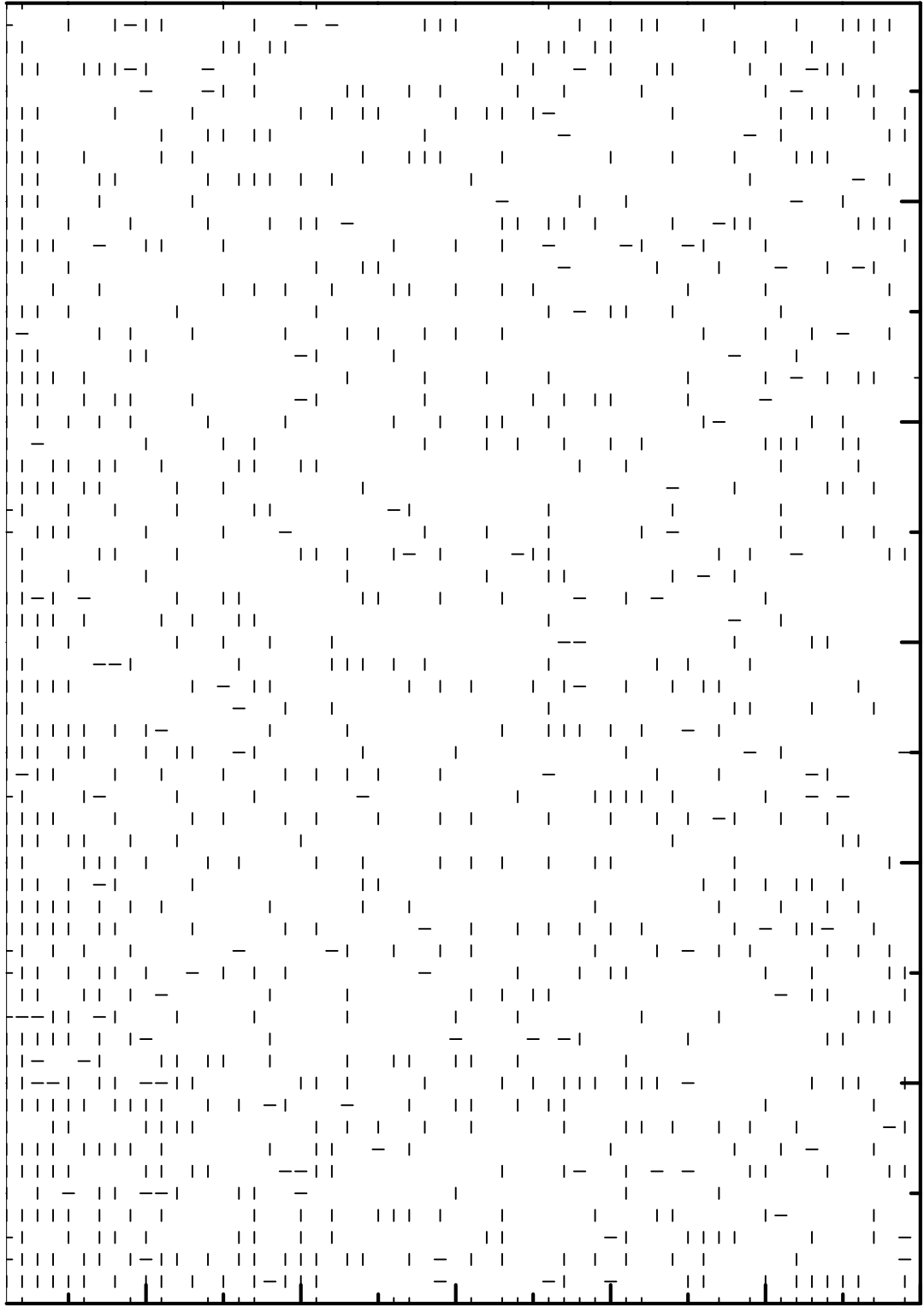
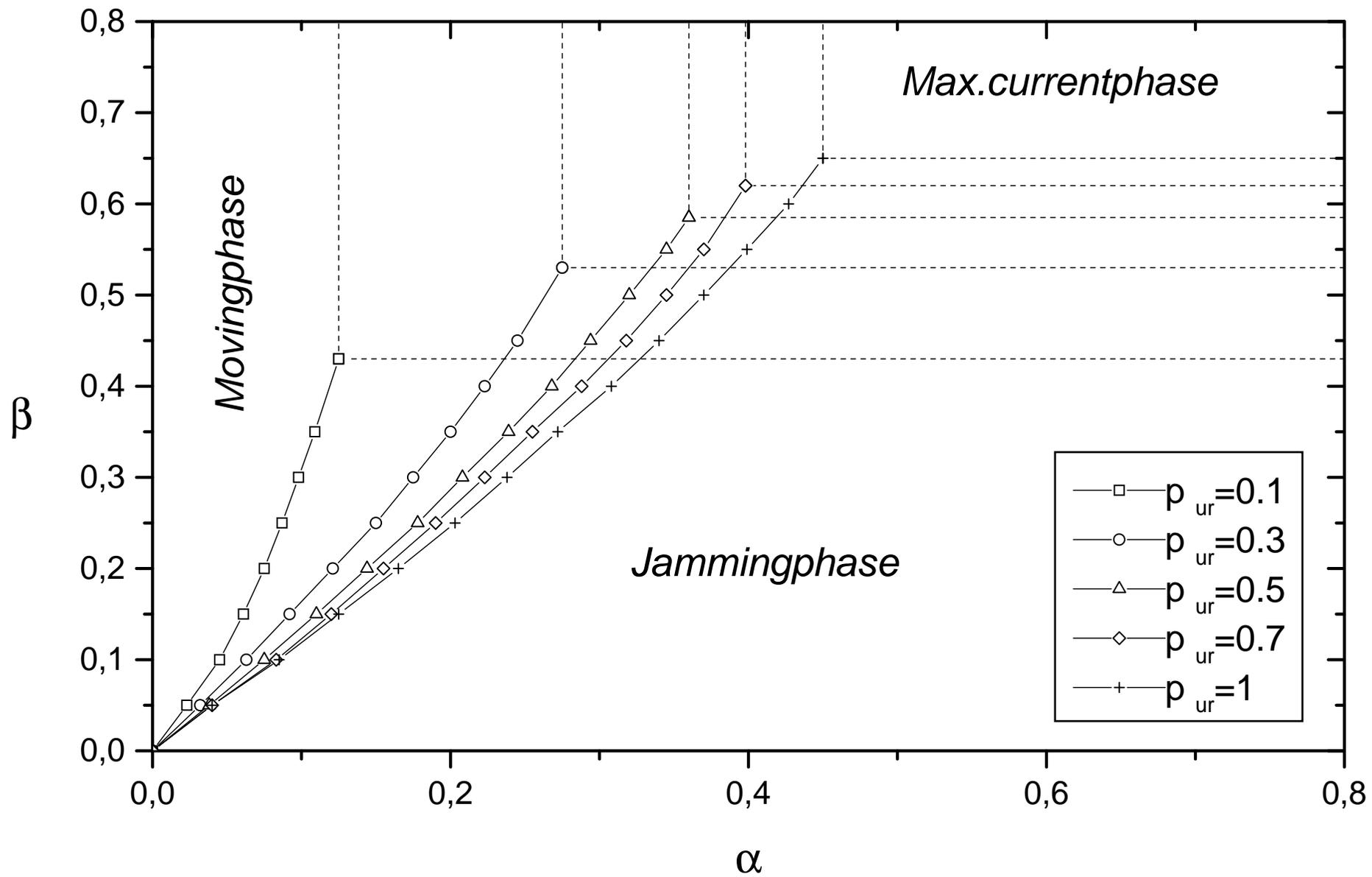
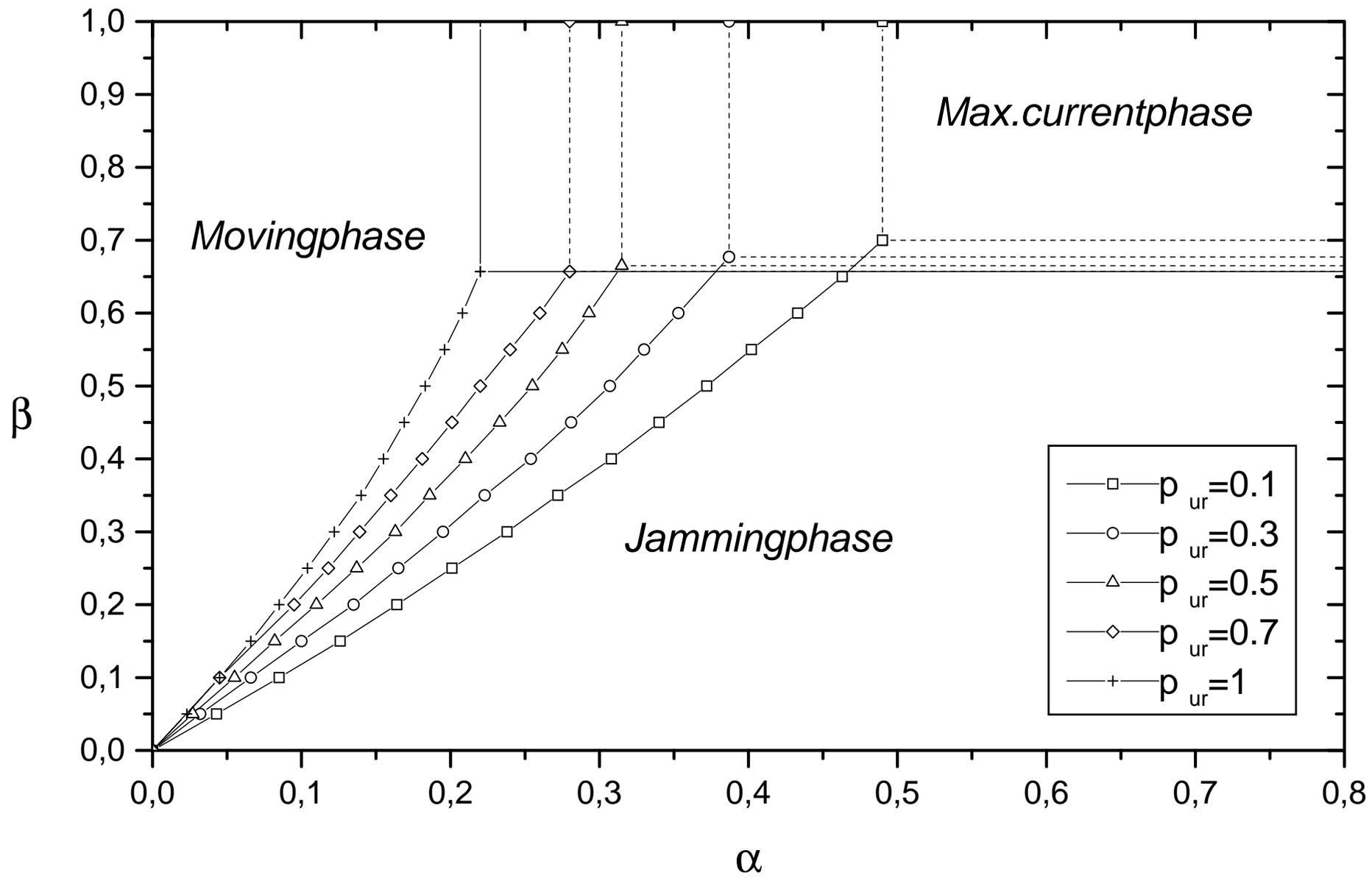
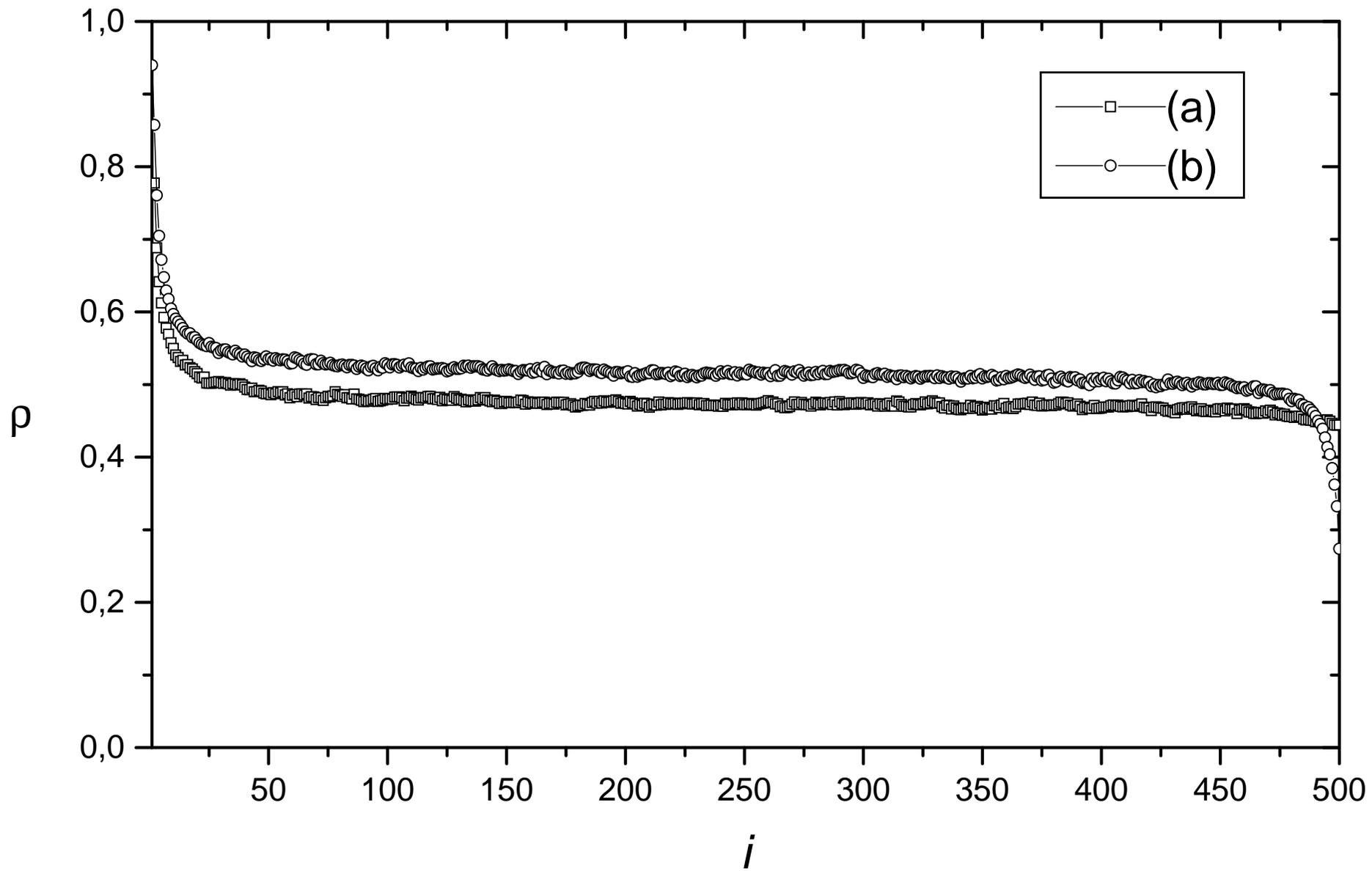
\end{document}